\documentclass{aa} 
\usepackage{amsmath} 
\usepackage{txfonts}
\usepackage{graphicx} 
\usepackage{url}
\usepackage{natbib} 
\usepackage{ulem}
\bibpunct{(}{)}{;}{a}{}{,}

\newcommand{\diff}{{\rm d}}

\newcommand{\Ms}{\ensuremath{M_\odot}}
\newcommand{\Zs}{$Z_\odot$}
\newcommand{\el}[2]{$\rm{}^{#2}\kern-0.6pt#1$}
\newcommand{\elm}[2]{\rm{}^{#2}\kern-0.8pt\rm{#1}}

\usepackage{color}
\definecolor{orange}{rgb}{1,0.5,0}

\usepackage{threeparttable}

\begin{document}

\title{Thermohaline instability and rotation-induced mixing}

\subtitle{III. Grid of stellar models and asymptotic asteroseismic quantities from the pre-main sequence up to the AGB for low- and intermediate-mass stars  at various metallicities}


\author{N. Lagarde\inst{1} \and T. Decressin\inst{1} \and C. Charbonnel\inst{1,2} \and P. Eggenberger\inst{1} \and S. Ekstr\"{o}m\inst{1} \and A. Palacios\inst{3}}

\institute{Geneva Observatory, University of Geneva, Chemin des
  Maillettes 51, 1290 Versoix, Switzerland \and
              IRAP, UMR 5277 CNRS and Universit\'e de Toulouse,  14, Av. E.Belin, 31400 Toulouse, France \and
              LUPM, Universit\'e Montpellier II, CNRS, UMR 5299,  Place E. Bataillon, 34095, Montpellier, France\ 
}

\date{Received / Accepted}

\authorrunning{} \titlerunning{III.Grid of stellar models and asteroseismic quantities}

\abstract
{The availability of asteroseismic constraints for a large sample of stars from the missions CoRoT and \textit{Kepler} paves the way for various statistical studies of the seismic properties of stellar populations.}
{In this paper, we evaluate the impact of rotation-induced mixing and
thermohaline instability on the global asteroseismic parameters at
different stages of the stellar evolution from the Zero Age Main
Sequence to theThermally Pulsating Asymptotic Giant Branch to distinguish stellar populations.}
{We present a grid of stellar evolutionary models for four metallicities
($Z=0.0001$, $0.002$, $0.004$, and $0.014$) in the mass range between  0.85 to
6.0~\Ms. The models are computed either with standard prescriptions or
including both thermohaline convection and rotation-induced mixing. For
the whole grid we provide the usual stellar parameters (luminosity,
effective temperature, lifetimes, ... ), together with the global
seismic parameters, i.e. the large frequency separation and asymptotic relations, the frequency corresponding to the maximum oscillation power $\nu_{\mathrm{max}}$, the maximal amplitude $A_{\mathrm{max}}$, the asymptotic period spacing of g-modes, and different acoustic radii.}
{We discuss the signature of rotation-induced mixing on the global asteroseismic quantities, that can be detected observationally.
Thermohaline mixing whose effects can be identified by spectroscopic studies cannot be caracterized with the global seismic
parameters studied here. But it is not excluded that individual mode frequencies or other well chosen asteroseismic quantities might help constraining this mixing.}
{}


\maketitle

\section{Introduction}

In the recent years important efforts were devoted to improve our
understanding of the physics of low- and intermediate-mass stars, 
and in particular to explain the abundance anomalies they exhibit along 
their lifetime.
Rotation was shown to change their internal dynamics due to the transport of angular momentum and chemical species through the action of meridional
circulation and shear turbulence, combined possibly with other processes induced by internal gravity waves or magnetic fields  \citep[see e.g.][]{Zahn92,Zahnetal97,MaeZah98,TalCha98, Eggenbergeretal05,ChTa05,ChTa08}. 
Rotation-induced mixing results in variations of the stellar chemical properties that 
successfully explain many of the abundance patterns observed  at the surface of these stars 
\citep{Palacios03,ChTa08,Smiljanic10,ChaLag10}. 
Additionally thermohaline mixing driven by $^3$He-burning was proposed to be the dominant process that further modifies the photospheric composition of bright low-mass red giant stars 
(\citealt{ChaZah07a}; for references on the abundance anomalies at that evolution phase see \citet{ChaLag10} hereafter Paper I, and \citet{Lagarde11}, hereafter Paper II). During the thermal-pulse phase on the asymptotic giant branch (TP-AGB), thermohaline mixing was found to lead to lithium production \citep[Paper I ;][]{Stancliffe10}, accounting for Li abundances observed in oxygen-rich AGB variables of the Galactic disk \citep{UttLeb10}.
In summary and as discussed in Papers I  and II of this series \citep[see also][]{ChaZah07a}, the effects of both rotation-induced mixing and thermohaline instability as described presently do account very nicely for most of the  spectroscopic observations of low- and intermediate-mass stars at various metallicities.

This has crucial consequences for the chemical evolution of the Galaxy (Paper II, and Lagarde et al., in preparation), and should also be taken into account in the other topical astrophysical domains that use stellar models as input physics. 
This is particuliarly true at the moment where asteroseismic probes bloom  across the Hertzsprung-Russel (HR) diagram. Thanks to the recent development of dedicated satellites as Corot and \textit{Kepler}, 
the internal properties of stars both on the main sequence
\citep[e.g.][]{Michel08,Chaplin2010} and
the giant branches \citep[e.g.][]{DeRidder2009,Bedding2010} are revealed. 
Furthermore due to the high number of stars observed by these missions, statistical studies are
possible through the determination of global pulsation properties like the frequency of maximum
oscillation power and the large frequency separation \citep[see e.g.][]{Miglio2009,Chaplin2011b}.

In this broad context the aim of the present paper is to provide the relevant classical stellar parameters together with the global asteroseismic properties  of low- and intermediate-mass stars all along their evolution.
This is done from the pre-main sequence (along the Hayashi track)  to the early-asymptotic giant branch (and along the TP-AGB for selected cases) for the grid of  models computed in Papers I and II for four metallicities spanning the range between $Z=0.0001$ and $Z=0.014$ and with initial masses between 0.85~M$_{\odot}$~ and 6.0~M$_{\odot}$.  This grid contains models computed with rotation-induced mixing and thermohaline instability, along with standard models without mixing outside convective regions for comparison purposes.
The present work is a prerequisite to validate further the current theoretical prescriptions for the non-standard mechanisms before we test them through detailed seismic dissections of individual stars. 

Such a grid of stellar models at different metallicities is obviously a key tool for various
important astrophysical topics related to e.g., stellar evolution in clusters,
stellar nucleosynthesis, chemical evolution, etc. They are available  since a long time for standard stellar models \citep[e.g.][]{Schalleretal1992,FoCh97,Yi03,Cassisi06}, and have recently appeared in the literature for rotation-induced models (\citealp{Brott2011},
\citet{Ekstrom11}). However these latest studies focus more on the evolution of massive stars, and do not include thermohaline mixing nor study the TP-AGB phase for low- and intermediate-mass stars as do those we present here.

The paper is organized as follows. In Sect.~2 we describe the physical inputs of the stellar evolution models. The table content of our grids 
is presented in Sect.~3.  Sect.~4 includes a short discussion on the main properties of the models and a comparison with the solar metallicity models of \citet{Ekstrom11}. 
In Sect.~5 we present the asteroseismic parameters of models with and without rotation.  Finally our main results are summarized in Sect.~6. 

\section{Physical inputs}

The models are computed with the lagrangian implicit stellar evolution code 
\textsc{STAREVOL} (v3.00. See  \citealt{SiessDufour2000, Palacios03, Palacios06, Decressin09}).
In this section we summarize the main physical ingredients used for the present grid.

\subsection{Basic inputs}

The description of the stellar structure rests on the hydrostatic and the continuity equations, {and the} equations for energy conservation and energy transport. To solve this system, the following physical ingredients are required : 

\begin{itemize}

 \item  Nuclear reaction rates are needed to follow the chemical changes inside burning sites, and to determine the production of energy by the nuclear reaction, $\epsilon_{nuc}$, and the energy loss by neutrino, $\epsilon_{\nu}$. We follow stellar nucleosynthesis with a network including 185 nuclear reactions involving 54 stable and unstable species from $^{1}$H to $^{37}$Cl. Numerical tables for the nuclear reaction rates were generated from NACRE compilation \citep{ArnouldGoriely1999,AikawaArnould2005} with the NetGen web interface\footnote{\url{http://www.astro.ulb.ac.be/Netgen/form.html}}.

We mainly use reactions rates from NACRE  or from \citet{CauFow88} when NACRE rates are not available. For proton captures on elements higher then Ne we follow rates from  \citet{Illiadisetal01} otherwise from \citet{Bao01}. The following reactions rates are computed by:

\begin{itemize}
\item $^{3}$He(D,p)$^{4}$He \citep{Descouvemont04}
\item $^{3}$He$(\alpha\alpha,\gamma)^{12}$C \citep{Fynbo05}
\item $^8$B($\beta,\nu$)2$^{4}$He ; $^{13}$N($\beta,\nu)^{13}$C ; $^{22}$Na($\beta,\nu$)$^{22}$Ne ; $^{26}$Alm($\beta,\nu)^{26}$Mg ; $^{26}$Alg($\beta,\nu)^{26}$Mg \citep{Horiguchi96} 
\item $^{14}$C(p,$\gamma)^{15}$N \citep{Wiescheretal90}
\item $^{14}$C(p,n)$^{14}$N \citep{KoeObr89}
\item $^{14}$C($\alpha$,n)$^{17}$O ; $^{17}$O(n,$^{4}$He)$^{14}$C \citep{Schatz93}
\item $^{14}$C($\alpha,\gamma)^{18}$O \citep{Funk89}
\item $^{14}$N(n,p)$^{14}$C \citep{Koehler89}
\item $^{14}$N(p,$\gamma)^{15}$O \citep{Mukhamedzhanov03}
\item $^{17}$O(n,$\gamma)^{18}$O \citep{Wagoner69}
\item $^{22}$Ne(p,$\gamma)^{23}$Na \citep{Haleetal02}
\item $^{22}$Ne(n,$\gamma)^{23}$Na \citep{Beer02}
\item $^{22}$Na(n,$\gamma)^{23}$Na ; $^{23}$Na($\alpha$,p)$^{26}$Mg ; $^{25}$Mg($\alpha$,p)$^{28}$Si ; $^{26}$Mg($\alpha$,p)$^{29}$Si ; $^{27}$Al($\alpha$,p)$^{30}$Si \citep{HaFe52}
\item  $^{26}$Alm(n,$\gamma)^{27}$Al ; $^{26}$Alg(n,$\gamma)^{27}$Al  \citep{Woosley78}\footnote{$^{26}$Alm and $^{26}$Alg represent the radioactive nuclide $^{26}$Al in its two isomeric states.}
\end{itemize}

\item The screening factors are calculated with the formalism of \citet{Mitler77} for weak and intermediate screening conditions and  of \citet{Graboske73} for strong screening conditions.  \\


\item Opacities are required to compute the radiative gradient $\nabla_{rad}$ and the energy transport by radiative transfer. 
We generate opacity tables according to \citet{IglRog96} using the
      OPAL
      website\footnote{\url{http://adg.llnl.gov/Research/OPAL/opal.html}} for $T>8000K$ that account for C and O enrichments. 
At lower temperature ($T<8000K$), we use the atomic and molecular opacities given by  \citet{Ferguson05}. \\


\item The equation of state 
relates the temperature, pressure, and density and thus provides different thermodynamic quantities ($\nabla_{\mathrm{ad}}$, $c_{\mathrm{P}}$,...). In STAREVOL we follow the formalism developed by \citet{EggletonFaulkner1973} and extended by \citet{PolsTout1995} that is based on the principle of Helmholtz free energy minimization (see \citealt{Dufour1999}, and \citealt{SiessDufour2000} for detailed description and numerical implementation). It accounts for the non-ideal effects due to Coulomb interactions and pressure ionization. \\


\item The treatment of convection is needed to compute the temperature gradient inside a convective zone. It is based on classical mixing length formalism with $\alpha_{MLT}=1.6$, from solar-calibrated models without atomic diffusion nor rotation computed by Geneva models \citep[see][]{Ekstrom11}. 
We assume instantaneous convective mixing, except when hot-bottom burning occurs on the TP-AGB, which requires a time-dependent convective diffusion algorithm as developed in \cite{FoCh97}.
The boundary between convective and radiative layers is defined with the Schwarzschild criterion. An overshoot parameter $d_{\mathrm{over}}/H_{\mathrm{p}}$ is taken into account for the convective core. This parameter is set to 0.05 or to 0.10 respectively for stars with masses below or above 2.0~\Ms \footnote{For small-sized core its mass extent is not allowed to be larger than $d_{\mathrm{over}}$ times the core mass.}. \\


\item We use a grey atmosphere where the photosphere is defined as the layer for which the optical depth $\tau$ is between 0.005 and 10. 
We define the effective temperature and radius at the layer where $\tau=2/3$.
\\
 
 \item For mass loss we use \citet{Reimers75} formula (with
       $\eta_{R}=0.5$) from the ZAMS up to central helium exhaustion
\begin{equation}
\dot{M} = -3.98~10^{-13} \eta_{R} \frac{LR}{M} ~M_{\odot}\mathrm{.yr^{-1}}.
\end{equation}
On AGB we shift to the mass loss prescription by \citet{VaWo93}. \\

\end{itemize}

\subsection{Transport processes in radiative zones}

\subsubsection{Thermohaline mixing \label{descriptionthermohaline}}

Thermohaline instability develops 
along the Red Giant Branch (RGB) at the bump luminosity in low-mass stars and on the early-AGB in intermediate-mass stars, when the gradient of molecular weight becomes negative  ($\nabla_{\mu} = \frac{d\ln \mu}{d \ln P} < 0$) in the external wing of the thin hydrogen-burning shell surrounding the degenerate stellar core \citep{ChaZah07a,ChaZah07b,Siess09,Stancliffe09,ChaLag10}.
This inversion of molecular weight is created by the $ \elm{He}3(\elm{He}3,2{\rm p})\elm{He}4$ reaction \citep{Ulrich71,Eggleton06,Eggleton08}.

The present grid of models is computed using the prescription advocated by \citet{ChaZah07a} and Paper I, II. It is based on
\citet{Ulrich72} with an aspect ratio of instability fingers $\alpha = 6$, in agreement with laboratory experiments \citep{Krish03}.  It includes the correction for non-perfect gas (including radiation pressure, degeneracy)  in the diffusion coefficient for thermohaline mixing that writes: 

\begin{equation}
 D_ t =  {C_ t} \,  {K}  \left({\varphi \over \delta}\right){- \nabla_\mu \over (\nabla_{\rm ad} - \nabla)} \quad \hbox{for} \;  \nabla_\mu < 0, 
\label{dt}
\end{equation}
with $K$ the thermal diffusivity ; $\varphi = (\partial \ln\rho / \partial \ln\mu)_{P,T}$ ; $\delta=-(\partial \ln\rho/ \partial \ln \nu)_{P,\mu}$ ; and with the non-dimensional coefficient
\begin{equation}
{C_ t} = {8 \over 3} \pi^2 \alpha^2. 
\end{equation}

The value of  $\alpha$ in actual stellar conditions was recently questionned by the results of 2D and 3D hydrodynamical simulations of thermohaline convection that favour $\alpha$ close to unity \citep{Denissenkov10,DenissenkovMerryfield10,RosenblumGaraudetal11,Traxleretal11}.  However these simulations are still far from the stellar regime; therefore we decided to use in this Series the prescription described above since it successfully reproduces the abundance data for evolved stars of various masses and metallicities (see Papers I and II for a more detailed discussion).

\subsubsection{Rotation-induced mixing}

Pre-main sequence evolution along the Hayashi track is computed in a standard way (i.e., without accounting for rotation-induced mixing), and solid-body rotation is assumed on the Zero  Age Main Sequence (ZAMS). On the main sequence, the evolution of the internal angular momentum profile is accounted with the complete formalism developed by \citet {Zahn92}, \citet {MaeZah98}, and \citet{MaZa04} that takes into account advection by meridional circulation and diffusion by shear turbulence (see \citealt{Palacios03,Palacios06,Decressin09} for a description of the implementation in STAREVOL). We do not take into account the inhibitory effects of $\mu$ gradients in the treatement of rotation.

We assume solid-body rotation in the convective regions as done in
Paper I and Paper II, considering that the transport of angular
momentum is dominated by the large turbulence in these regions that
instantaneously flattens out the angular velocity profile as for the
abundance profiles. This hypothesis leads to a minimum shear mixing
approach in the underlying radiative layers as discussed in
\citet{Palacios06}, and \citet{BrunPalacios09}. \\
On the other hand the transport of angular momentum in stellar radiative layers obeys an
advection/diffusion equation : 
 
 \begin{equation}
 \rho\frac{\diff (r^{2}\Omega)}{\diff t}=\frac{1}{5r^{2}}\frac{\partial}{\partial r}(\rho r^{4} \Omega U_{r}) + \frac{1}{r}\frac{\partial}{\partial r}\left(r^{4}\rho \nu_{\nu}\frac{\partial \Omega}{\partial r}\right)
 \end{equation}
where $\rho, r$, and $\Omega$ have their usual meaning.  
$U_{r}$ is the vertical component of meridional circulation velocity, and $\nu_{\nu}$ is the vertical component of the turbulence viscosity. 

The transport of chemical species resulting from meridional circulation and both vertical and horizontal turbulence is computed as a diffusive process \citep{Chaboyer92}. The vertical transport of a chemical species i of concentration $c_{i}$ is described by a pure diffusion equation :

\begin{equation}
 \frac{\diff c_{i}}{\diff
  t}=\underbrace{\dot{c_{i}}}_\text{nuclear}+\underbrace{\frac{1}{\rho r^{2}}\frac{\partial}{\partial
    r}\left(r^{2}\rho D_\text{tot} \frac{\partial c_{i}}{\partial
      r}\right)}_\text{diffusion processes}
\end{equation}
where $\dot{c_{i}}$ represents the variations of chemical composition due to nuclear reactions. The total diffusion coefficient $D_\text{tot}$ for chemicals can be written as the sum of three coefficients: 

\begin{equation}
D_\text{tot}=D_\text{th}+D_\text{eff}+D_\text{v}
\end{equation}
with $D_\text{th}$ the thermohaline coefficient
(Sect.~\ref{descriptionthermohaline}), $D_\text{eff}$ the effective
diffusion coefficient ($\propto 1/D_\text{h}$), and $D_\text{v}$ the
vertical turbulent diffusion coefficient \citep{TalZah97}  that is
proportional to the
horizontal diffusion coefficient \citealt{Chaboyer92}). The expression
of the horizontal diffusion coefficient is taken from \cite{Zahn92},
with its expression that prevents numerical diverge :

\begin{equation}
D_{h}=\frac{r}{C_{h}}(|2V-\alpha U|^2+U^2)^{1/2}
\end{equation}

 We do not consider possible interactions between thermohaline and rotation-induced mixing, nor magnetic diffusion. 
Under the present assumptions the thermohaline diffusion coefficient is several orders of magnitude higher than the total diffusion coefficient related to rotation and than magnetic diffusivity in the advanced phases where the thermohaline instability develops (see e.g., \citealt{ChaLag10,CantielloLanger2010}, and Paper II). 

The complete treatment is aplied up to the RGB tip or up to the second dredge-up for star undergoing the He-flash episode.

\subsection{Initial rotation velocity}

The initial rotation velocity of our models on the ZAMS is chosen at $45\%$ of the critical velocity at that point, with $V_\text{crit}=\left(\frac{2}{3}\right)^{\frac{3}{2}} \left(\frac{GM}{R}\right)^{\frac{1}{2}}$.
Here we take R the stellar radius computed without considering the stellar deformation due to rotation ; if we were to take into account the deformation of the stellar radius as in \citet{Ekstrom11}, then  our initial velocities would correspond to $30\%$ of critical velocity.  
This choice of $V_\text{ini}/V_\text{crit} = 0.45$ fits well the mean value in the observed velocity distribution of low- and intermediate-mass stars in young open clusters. 
This initial rotation rate leads to mean velocities on the main sequence between 90 and 137 km.s$^{-1}$.\\

We apply magnetic braking only to [1.25~M$_{\odot}$~; Z$_{\odot}$] and [0.85~M$_{\odot}$~ ;
$Z=0.0001$] following the description of \citet{Kawaler88} according to \citet {TalCha98} and \citet
{ChaTal99}.

The transport by internal gravity waves, which is efficient only in main sequence stars with effective temperatures on the ZAMS lower than 6500 K \citep[see][]{TalCha03}, is neglected. We do not account either for dynamo processes nor for the presence of fossil magnetic fields.

\subsection{Initial abundances}

\begin{table}[t]
  \caption{Initial abundances in mass fraction for the models at different metallicities}
  \begin{tabular}{rcccc}
    \hline
    {[Fe/H]} & 0 & -0.56 & -0.86 & -2.16 \\ 
    Element & Z = 0.014 & Z = 0.004 & Z = 0.002 & Z = 0.0001 \\
    \hline
    \hline
    \el{H}{1}   & $7.20\cdot10^{-01}$ & $7.42\cdot10^{-01}$   & $7.47\cdot10^{-01}$   & $7.52\cdot10^{-01}$  \\
    \el{H}{2}   & $ 3.74\cdot10^{-05}$   & $3.86\cdot10^{-05 }$ & $3.89\cdot10^{-05}$  & $3.91\cdot10^{-05 }$\\
    \el{He}{3}  &$ 2.83\cdot10^{-05}$   & $2.69\cdot10^{-05 }$ &$ 2.67\cdot10^{-05 }$ & $2.64\cdot10^{-05 }$\\
    \el{He}{4}  & $2.66\cdot10^{-01 }$  &$ 2.53\cdot10^{-01}$  &$ 2.50\cdot10^{-01 }$ &$ 2.48\cdot10^{-01}$ \\
    \el{Li}{6}  & $6.35\cdot10^{-10}$  &$ 1.81\cdot10^{-10}$ &$ 9.07\cdot10^{-11}$ & $4.53\cdot10^{-12}$ \\
    \el{Li}{7}  & $ 9.00\cdot10^{-09}$  & $2.16\cdot10^{-09}$  & $2.17\cdot10^{-09}$ & $2.18\cdot10^{-09}$ \\
    \el{Be}{9}  & $ 1.69\cdot10^{-10}$  & $4.82\cdot10^{-11}$  & $2.41\cdot10^{-11}$ & $1.21\cdot10^{-12}$ \\
    \el{B}{10}  & $ 8.09\cdot10^{-10}$  & $2.31\cdot10^{-10}$ & $1.16\cdot10^{-10}$ & $5.78\cdot10^{-12}$ \\
    \el{B}{11}  & $ 3.94\cdot10^{-09 }$  & $1.13\cdot10^{-09}$  & $5.63\cdot10^{-10}$ & $2.82\cdot10^{-11}$ \\
    \el{C}{12}  & $ 2.27\cdot10^{-03 }$  & $6.47\cdot10^{-04}$  & $3.24\cdot10^{-04}$  & $1.62\cdot10^{-05}$ \\
    \el{C}{13}  & $ 3.63\cdot10^{-05 }$  & $1.04\cdot10^{-05}$  & $5.19\cdot10^{-06}$  & $2.59\cdot10^{-07}$ \\
    \el{N}{14}  & $6.56\cdot10^{-04 }$  & $1.87\cdot10^{-04}$  & $9.38\cdot10^{-05}$  & $4.69\cdot10^{-06}$ \\
    \el{N}{15}  & $2.34\cdot10^{-06 }$  & $6.69\cdot10^{-07}$  & $3.35\cdot10^{-07}$  & $1.67\cdot10^{-08}$ \\
    \el{O}{16}  & $5.69\cdot10^{-03 }$  & $1.62\cdot10^{-03}$  & $8.13\cdot10^{-04}$  & $4.06\cdot10^{-05}$ \\
    \el{O}{17}  & $3.82\cdot10^{-06 }$  & $1.09\cdot10^{-06}$  & $5.46\cdot10^{-07}$  & $2.73\cdot10^{-08}$ \\
    \el{O}{18}  & $1.28\cdot10^{-05}$ & $3.67\cdot10^{-06}$  & $1.83\cdot10^{-06}$  & $9.17\cdot10^{-08}$ \\
    \el{F}{19}  & $5.38\cdot10^{-07 }$  & $1.54\cdot10^{-07}$  & $7.69\cdot10^{-08}$  & $3.85\cdot10^{-09}$ \\
    \el{Ne}{20} & $1.79\cdot10^{-03 }$  & $5.10\cdot10^{-04}$  & $2.55\cdot10^{-04}$  & $1.28\cdot10^{-05}$ \\
    \el{Ne}{21} & $5.70\cdot10^{-06 }$  & $1.63\cdot10^{-06}$  & $8.14\cdot10^{-07}$  & $4.07\cdot10^{-08}$ \\
    \el{Ne}{22} & $2.40\cdot10^{-04 }$  & $6.85\cdot10^{-05 }$ & $3.42\cdot10^{-05}$  & $1.71\cdot10^{-06}$ \\
    \el{Na}{23} & $2.65\cdot10^{-05 }$  & $7.58\cdot10^{-06}$  & $3.79\cdot10^{-06}$  & $1.89\cdot10^{-07}$ \\
    \el{Mg}{24} & $4.99\cdot10^{-04 }$  & $1.42\cdot10^{-04}$  & $7.13\cdot10^{-05}$  & $3.57\cdot10^{-06}$ \\
    \el{Mg}{25} & $6.69\cdot10^{-05 }$  & $1.91\cdot10^{-05}$  & $9.56\cdot10^{-06}$  & $4.78\cdot10^{-07}$ \\
    \el{Mg}{26} & $7.67\cdot10^{-05 }$  & $2.19\cdot10^{-05}$  & $1.10\cdot10^{-05}$  & $5.48\cdot10^{-07}$ \\
    \el{Al}{27} & $4.94\cdot10^{-05 }$  & $1.41\cdot10^{-05}$  & $7.05\cdot10^{-06}$  & $3.52\cdot10^{-07}$ \\
    \el{Si}{28} & $5.97\cdot10^{-04 }$  & $1.71\cdot10^{-04}$  & $8.53\cdot10^{-05}$  & $4.27\cdot10^{-06 }$\\
    \el{Si}{29} & $6.65\cdot10^{-05 }$  & $1.90\cdot10^{-05}$  & $9.50\cdot10^{-06}$  & $4.75\cdot10^{-07 }$\\
    \el{Si}{30} & $4.81\cdot10^{-05 }$  & $1.38\cdot10^{-05}$  & $6.87\cdot10^{-06}$  & $3.44\cdot10^{-07}$ \\
    \el{P}{31}  & $5.54\cdot10^{-06 }$  & $1.58\cdot10^{-06}$  & $7.91\cdot10^{-07}$  & $3.96\cdot10^{-08}$ \\
    \el{S}{32}  & $3.24\cdot10^{-04 }$  & $9.26\cdot10^{-05}$  & $4.63\cdot10^{-05}$  & $2.31\cdot10^{-06}$ \\
    \el{S}{33}  & $3.48\cdot10^{-06 }$  & $9.95\cdot10^{-07}$  & $4.97\cdot10^{-07}$  & $2.49\cdot10^{-08}$ \\
    \el{S}{34}  & $1.80\cdot10^{-05 }$  & $5.15\cdot10^{-06}$  & $2.58\cdot10^{-06}$  & $1.29\cdot10^{-07}$ \\
    \el{Cl}{35} & $6.54\cdot10^{-06 }$  & $1.87\cdot10^{-06}$  & $9.34\cdot10^{-07}$  & $4.67\cdot10^{-08}$ \\
    \el{Cl}{37} & $2.15\cdot10^{-06 }$  & $6.14\cdot10^{-07}$  & $3.07\cdot10^{-07}$  & $1.53\cdot10^{-08}$ \\
    Others      & $1.47\cdot10^{-03 }$  & $4.19\cdot10^{-04}$  & $2.09\cdot10^{-04}$  & $1.05\cdot10^{-05}$ \\
    \hline
  \end{tabular}
  \label{tab:initab}
\end{table}

Table \ref{tab:initab} presents the initial abundances in mass fraction that we assume at different metallicities.
We adopt the solar mixture of  \citet{Asplund09}, except for Ne for which we use value derived by \citet{Cunha06}.
We use the ratio $\Delta Y / \Delta Z =1.29$ derived by \citet{Ekstrom11} to account for the enrichment in helium reported  to enrichment in heavy elements in the Galaxy until the birth of the Sun. For the primordial abundances we take the 
 WMAP-SBBN value from \citet{Coc04}. 
 To determine the initial composition of our models at a given metallicity Z we use the scaling  ($X_{i}=X_{i,\odot}\cdot\frac{Z}{Z_{\odot}}$) for all elements except for $^{7}Li$ which is taken constant (Li/H = 4.15.10$^{-10}$) for $\frac{Z}{X}<7.8.10^{-3}$.
We assume $[\alpha/\text{Fe}]=0$ at all metallicities, which has a negligible impact in the present context\footnote{Using $[\alpha/\text{Fe}]=+0.3$ instead of 0 for our [1.5~M$_{\odot}$; $Z=0.0001$] model leads to a decrease of the main-sequence lifetime by $\sim$2\%, and lowers the 
turnoff luminosity and effective temperature by  5\% and 1\% respectively. This is negligible compared to the effects of rotation.}.

\section{Description of grids}

\subsection{Content of electronic tables \label{tablecontent}}

\begin{figure}
  \includegraphics[angle=0,width=8.5cm,clip=true, trim= 0cm 0cm 0cm 1cm]{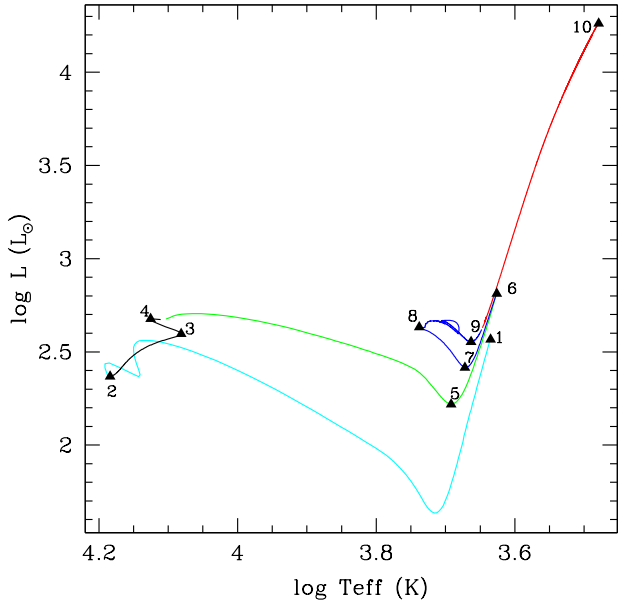}
  \caption{Evolution track in the Hertzsprung-Russell diagram  of the standard 4.0~\Ms~ model at solar metallicity. Each phase is indicated by a different color : pre-main sequence (cyan), main sequence (black), Hertzsprung gap and red giant branch (green), core helium-burning phase (blue), and asymptotic giant branch (red). Black triangles show the points that we have selected to describe the tracks (see Sect.\ref{tablecontent})}
  \label{fig:Hr_ptsgrilles}
\end{figure}

We provide files containing sets of relevant physical
quantities as a function of time that characterize our models computed in the initial mass range between 0.85~M$_{\odot}$~ and 6.0~M$_{\odot}$~ with four metallicities
$Z=0.0001$, 0.002, 0.004 and 0.014 ($\text{[Fe/H]= -2.16}$, -0.86, -0.56,
and 0 respectively). For each mass and metallicity, models are computed
with standard prescriptions (no mixing other than convection process),
and with both thermohaline instability and rotational transport. For all masses the evolution is followed from the beginning of the pre-main sequence (along the Hayashi track) up to the early-AGB phase.  
For each model, we have selected 500 points to allow a good description of the full raw tracks. 
First, key evolutionary points are determined (see Fig.~\ref{fig:Hr_ptsgrilles}): 
\begin{enumerate}
\item beginning of the pre-main sequence;
\item ZAMS  defined as the time when the central
  hydrogen abundance $X_c$ has decreased by 0.003 in mass fraction compared to its initial value.
\item turning point with the lowest $T_\text{eff}$ on the main sequence;
\item end of core H-burning defined as the point beyond which $X_c$ is
      smaller than $10^{-7}$ ; 
\item bottom of the Red Giant Branch (RGB);
\item RGB-tip;
\item local minimum of luminosity during central He-burning;
\item local maximum of $T_\text{eff}$ during central He-burning;
\item bottom of AGB : point with local minimum of luminosity after the loop on HB ;
\item end of core He-burning defined as the point beyond which $Y_c$ is
      smaller than $10^{-4}$. 
\end{enumerate}

\begin{table*}[htbf]
\hspace{3cm}
  \caption{Description of table containing result of our evolution models, both standard case and including rotation and thermohaline mixing.}
   \centering 
  \scalebox{0.92}{
  \begin{tabular}{ l  l  l }
  \hline \hline
  \bf{Stellar parameters}                                            & \bf{Surface abundances}                       & \bf{Central abundances}                          \\
  \hline \hline
 - Model number                                                                       &  $^{1}$H $^{2}$H      &   $^{1}$H          \\
 - Maximum of temperature $T_{\mathrm{max}}$	 (K)   &              $^{3}$He $^{4}$He                                               & $^{3}$He $^{4}$He                \\
 - Mass coordinate of $T_{\mathrm{max}}$    (\Ms)     &            $^{6}$Li $^{7}$Li                                                    &                      \\
 - Effective temperature T$_{\mathrm{eff}}$ (K)                      &     $^{7}$Be $^{9}$Be                        &                              \\
 - Surface luminosity L (L$_{\odot}$)                                       &     $^{10}$B $^{11}$B                     &                  \\
 - Photospheric radius radius R$_{eff}$ (R$_{\odot}$)                   &$^{12}$C $^{13}$C $^{14}$C                             &   $^{12}$C $^{13}$C $^{14}$C                   \\
 - Photospheric density $\rho_{eff}$ (g.$\mathrm{cm^{-3}}$)       &        $^{14}$N $^{15}$N                     &            $^{14}$N           \\
 - Density at the location of $T_{max}$, $\rho_{max}$ (g.$\mathrm{cm^{-3}}$)        &$^{16}$O  $^{17}$O $^{18}$O                          &       $^{16}$O  $^{17}$O $^{18}$O                \\
 - Stellar mass M (\Ms)                                  &  $^{19}$F                              &         $^{19}$F       \\
 - Mass loss rate ($\mathrm{\Ms.yr^{-1}}$)                          &$^{20}$Ne $^{21}$Ne $^{22}$Ne   &     $^{20}$Ne $^{21}$Ne $^{22}$Ne                 \\
 - Age t (yr)                                                                          &  $^{23}$Na     &      $^{23}$Na      \\
 - Photospheric gravity $log (g_{\mathrm{eff}})$  (log(cgs))  &  $^{24}$Mg $^{25}$Mg $^{26}$Mg                         &     $^{24}$Mg $^{25}$Mg $^{26}$Mg             \\
 - Central temperature $T_{c}$ (K)    &  $^{26}$Al $^{27}$Al       &     $^{26}$Al $^{27}$Al     \\
 - Central pressure $P_{c}$     &  $^{28}$Si                                            &     $^{28}$Si              \\
 - Surface velocity $v_{surf}$ ($\mathrm{km.s^{-1}}$) &&\\ 
 &&\\
 - Mass at the base of convective envelope (\Ms)        &&\\
                                                                                      &&\\
 - The large separation from scaling relation $\Delta \nu_{\mathrm{scale}}$ ($\mathrm{\nu Hz}$) & & \\
 - The large separation from asymptotic relation $\Delta \nu_{\mathrm{asymp}}$ ($\mathrm{\nu Hz}$) & & \\
 - The frequency with the maximum amplitude $\nu_{\mathrm{max}}$& &\\
 - The maximum amplitude $A_{\mathrm{max}}$ & & \\ 
 - The asymptotic period spacing of g-modes $\Delta\mathrm{\Pi}$ (s)& & \\ 
 - The total acoustic radius T (s) & & \\  
 - The acoustic radius at the base of convective evelope t$_{\mathrm{BCE}}$ (s)  & & \\  
 - The acoustic radius at the location of helium second-ionisation region t$_{\mathrm{He}}$ (s)& & \\

  \hline \hline  
  \end{tabular}}
  \label{tab:gridtable}
\end{table*}

The data in the tables are linearly interpolated within the results of
each evolutionary model. We perform the interpolation as a function of time, central
mass fraction of hydrogen or helium, or luminosity according to the
evolutionary phase, and the final interpolated values are given at 499 points that we distribute
as follows:
\begin{itemize}
\item 99 points evenly distributed in time sample the pre-main
      sequence between points 1 and 2.
\item  110
      points evenly distributed in terms of the central hydrogen mass fraction
      $X_{C}$ sample the main sequence with 85 points distributed
      between points 2 and 3, and 25 points distributed between points 3
      and 4.
\item 60 points evenly spaced in time sample the
      Herzsprung gap between points 4 and 5. 
\item 80 points evenly distributed in
      terms of  $\log L$ sample the RGB between points 5 and 6.
\item 20, 70 and 70 points evenly distributed in terms of $Y_C$
      respectively sample the central He burning phase between
points 6 and 7\footnote{For low-mass stars below 2.0~\Ms{} in which He-flash
  occurs we do not include table points between evolutionary points 6 and
  7.} , 7 and 8 and,  8 and 9.
\item Finally, 50 points evenly spaced in $\log L$ are selected between points 9 and
10  
\end{itemize}

For each model, we store the quantities given in table
\ref{tab:gridtable} in a file that can be 
retrieved from \url{http://obswww.unige.ch/Recherche/evol/-Database-} .

\begin{figure}[t]
  \includegraphics[width=0.25\textwidth,trim= 0cm 0cm 9cm 0cm]{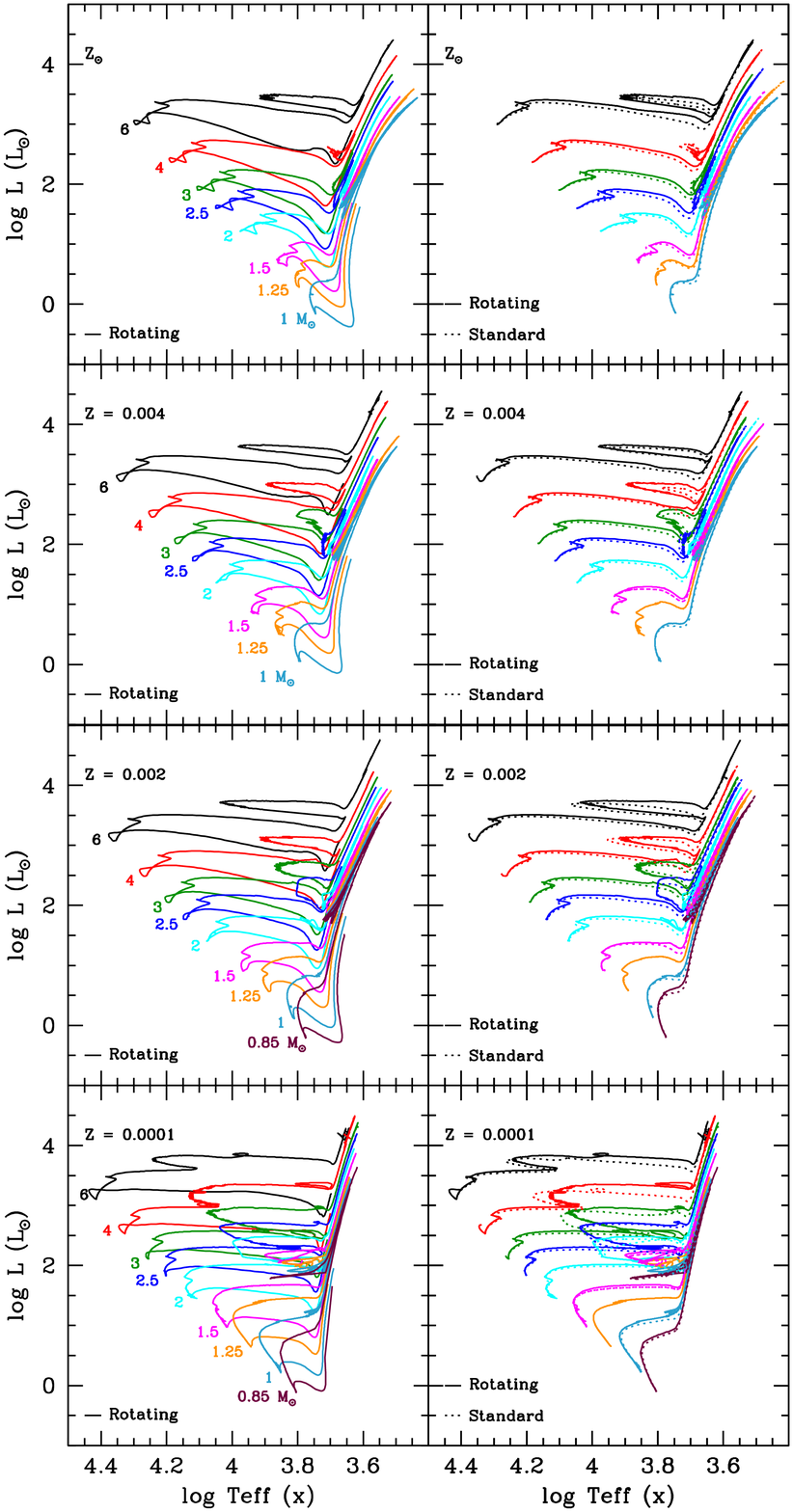}
  \caption{Theoretical evolution tracks in the HR diagram for the ensemble of calculated models for all metallicities ($Z_\odot$, Z=0.004, Z=0.002, and Z=0.0001, from top to bottom). 
  On left panels the tracks are shown for the rotating models from the pre-main sequence on.
  On the right panels, both standard and rotating predictions are shown
 (dashed and solid lines respectively) from the ZAMS and beyond.}
  \label{fig:HRD}
\end{figure}

\subsection{Comparisons between models from \citet[][Geneva code]{Ekstrom11} and our models (STAREVOL)}

\citet{Ekstrom11} computed a large grid of models with rotation from 0.8 to 120~M$_{\odot}$ at solar metallicity with the Geneva stellar evolution code. The present grid is complementary since  the low-mass star models of \citet{Ekstrom11} are computed only to the helium flash at the RGB tip and do not include thermohaline mixing. Also, we presently explore asterosismic diagnostics during the TP-AGB.

In order to offer this complementarity between the 2 sets of giants, we were careful in choosing the same input physics and assumptions, although some differences remain that we describe below. 
Both codes use the same physical inputs for convection (Schwarzschild criterion and overshoot), opacities, mass loss, and nuclear reaction rates, and for the mass domain explored the  different equations of state have a negligible impact on the stellar structures. 
Besides, the initial abundances of our models are similar apart from the larger number of species followed in STAREVOL.  STAREVOL has indeed a more extended network of nuclear reactions than the Geneva code, which allows to follow in particular the evolution of unstable elements such as $^{13}N$, $^{15}O$, and $^{7}Be$. As a consequence, the convective cores are smaller on the main sequence in STAREVOL models (by $\sim$7\% for the [4.0 M$_{\odot}$, Z$_{\odot}$] model), and the tracks in the HR-diagram are slightly less luminous ($L_\text{TO}$ is $\sim$10\% lower for the  [4.0 M$_{\odot}$, Z$_{\odot}$] model). This results in a difference in the lifetime on the main sequence 4\% for the [4.0 M$_{\odot}$, Z$_{\odot}$] model. This is the main difference we could identify when comparing standard models computed with the two codes. \\

For the rotational transport both codes follow the advection by meridional circulation and the diffusion by shear turbulence. We carefully checked that the use of a different prescription for the turbulent diffusion coefficient $D_{v}$ (\citet{TalZah97} in STAREVOL and \citet{Maeder97} in Geneva code) has no impact in the mass domain we explore as the mixing above the convective core is dominated by $D_\text{eff}$. 
Both codes use prescriptions from \citet{Zahn92} for the horizontal diffusion coefficient, $D_{h}$. However in STAREVOL we always use Eq.7 that is given by \citet{Zahn92} to avoid the divergence of effective diffusion coefficient ($\propto$ $\frac{1}{D_{h}}$), while in Geneva the original Eq.2.29 of \citet{Zahn92} is used. Surprisingly Eq.7 implies higher values of $D_{h}$ at the edge of the convective core (by about a factor of 4 in the [4.0 M$_{\odot}$, Z$_{\odot}$] model on the main sequence), which leads to less mixing in these central regions. 
Combined with the fact that STAREVOL models have a lower initial rotation velocity on the zero age main sequence, this results in a lower lifetime and lower luminosity (the maximum difference being of 15\% for the [4.0 M$_{\odot}$, Z$_{\odot}$] model).
  
\section{Global properties of the models \label{globalppt}}

The effects of rotation-induced mixing and thermohaline instability on
the surface chemical properties of the grid stars were extensively
discussed in Papers I and II. Here we focus on the global properties of
the models

\subsection{Hertzsprung-Russell diagrams and logTc vs log$\rho_c$}
The evolutionary tracks in the HR diagram are shown in  Fig.\ref{fig:HRD} for all the models computed in this study, with and without rotation. We also present the evolution of the central temperature and density in Fig.\ref{fig:rhoT} for the solar-metallicity case.

\subsubsection{Metallicity effects} \label{Zeffects}

Let us first recall the impact of metallicity as  already reported in the literature \citep[e.g.][]{Schalleretal1992, Schaerer93, Charbonnel94,HeLa00,Maeder09}. For a given stellar mass one notes the following effects when metallicity decreases, due to the lowering of the radiative opacity:

\begin{itemize}

\item The ZAMS is  shifted to the
      blue, and both the stellar luminosity and effective temperature
      are higher at that phase compared to solar metalicity models.
\item The positions of the RGB and of the AGB are shifted to the blue, and the  ignition of helium (in degenerate conditions or not) occurs at lower luminosity at the tip of the RGB (L=21308 L$_{\odot}$ and 21252 L$_{\odot}$ for models [1.5M$_{\odot}$, Z=0.014 and Z=0.002] respectively).
\item Central helium burning occurs at a higher effective temperature,
      and more extended blue loops are obtained.
\end{itemize}

\subsubsection{Impact of rotation-induced mixing and thermohaline instability}

\begin{figure*}[t]
  \includegraphics[width=0.5\textwidth]{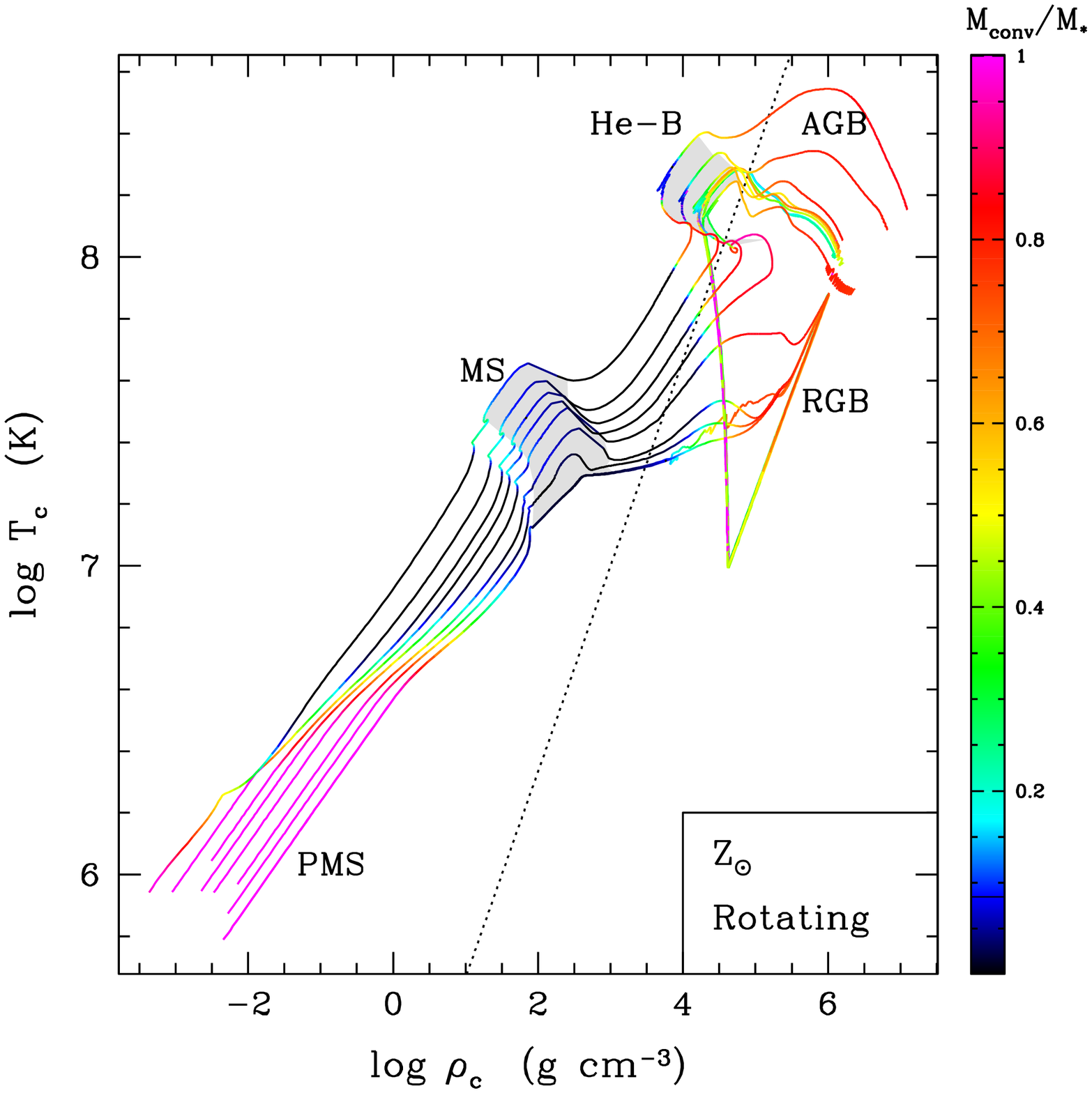}
  \includegraphics[width=0.5\textwidth]{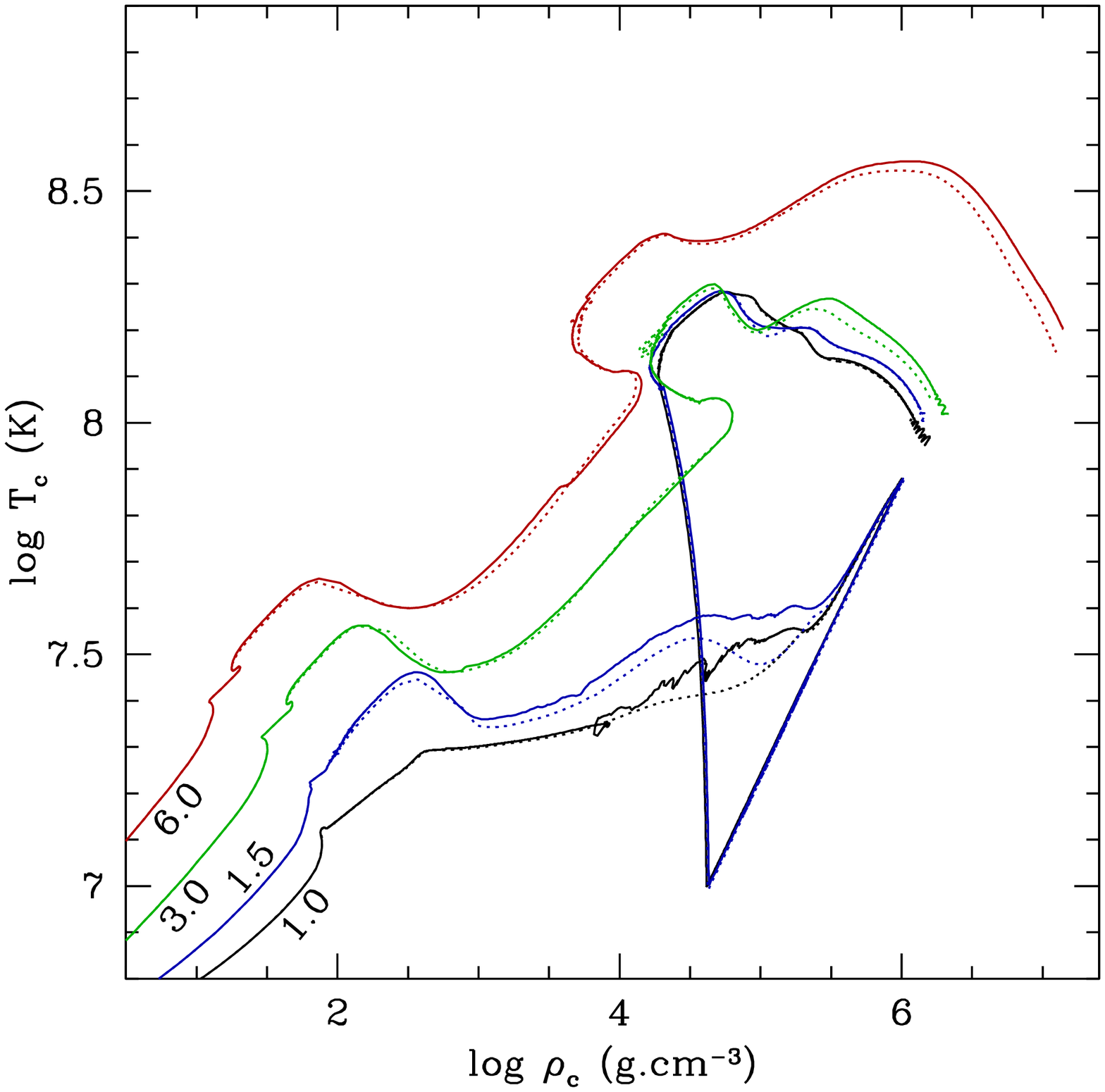}
  \caption{{\sl Left panel} Central density as a function of central
 temperature for the rotating \Zs{} models. Colours indicate the mass of
 convective regions (convective core and convective envelope) over the
 total stellar mass. Shaded regions indicate the phases of central
 hydrogen-burning (MS) and helium-burning (He-B). {\sl Right panel} Central
 density as a function of central temperature for standard and rotating
 models (dashed and solid lines respectively) for four initial stellar
 masses from the ZAMS beyond. 
  }
  \label{fig:rhoT}
\end{figure*}

\begin{itemize}

\item As known for a long time \citep[see e.g.][]{MaMe00} and as shown in Fig.\ref{fig:HRD}, rotation-induced mixing affects the evolution tracks in HR diagram.
On the main sequence, rotational mixing brings fresh hydrogen fuel into the longer lasting convective core, and transports H-burning products outwards. This results in more massive helium cores at the turnoff than in the standard case and shifts the tracks towards higher effective temperatures and luminosities all along the evolution  \citep{Ekstrom11,HeLa00,MaMe00,MeMa00}. 
The right panel of Fig.\ref{fig:rhoT} shows the central conditions in standard and rotating models for selected masses at solar metallicity. Rotating models behave as star with higher mass during all of their evolution.  

\item  Thermohaline mixing induced by $^3$He-burning becomes efficient only on the RGB at the bump luminosity \citep{ChaZah07a,ChaLag10}.  
Beyond this point the double-diffusive
       instability develops in a very thin region located between the
       hydrogen-burning shell and the convective envelope, and has
       negligible effect on the stellar structure. It does not modify the  evolution tracks in the HR diagram nor in the log T$_c$ vs log~$\rho_C$ diagram.

\end{itemize}
 
\subsection{Lifetimes}

\begin{figure*}[t]
  \includegraphics[width=0.5\textwidth]{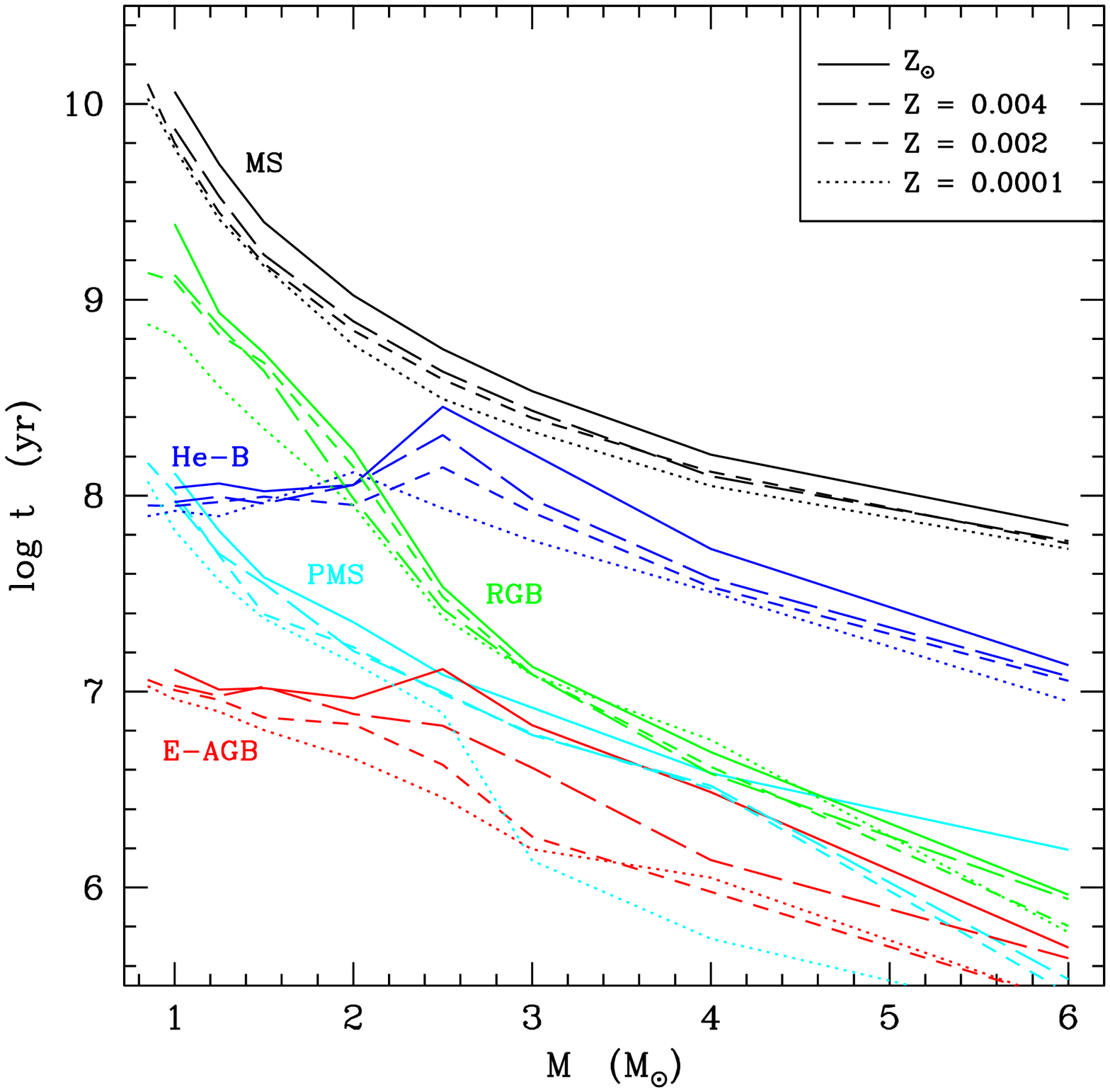}
  \includegraphics[width=0.5\textwidth]{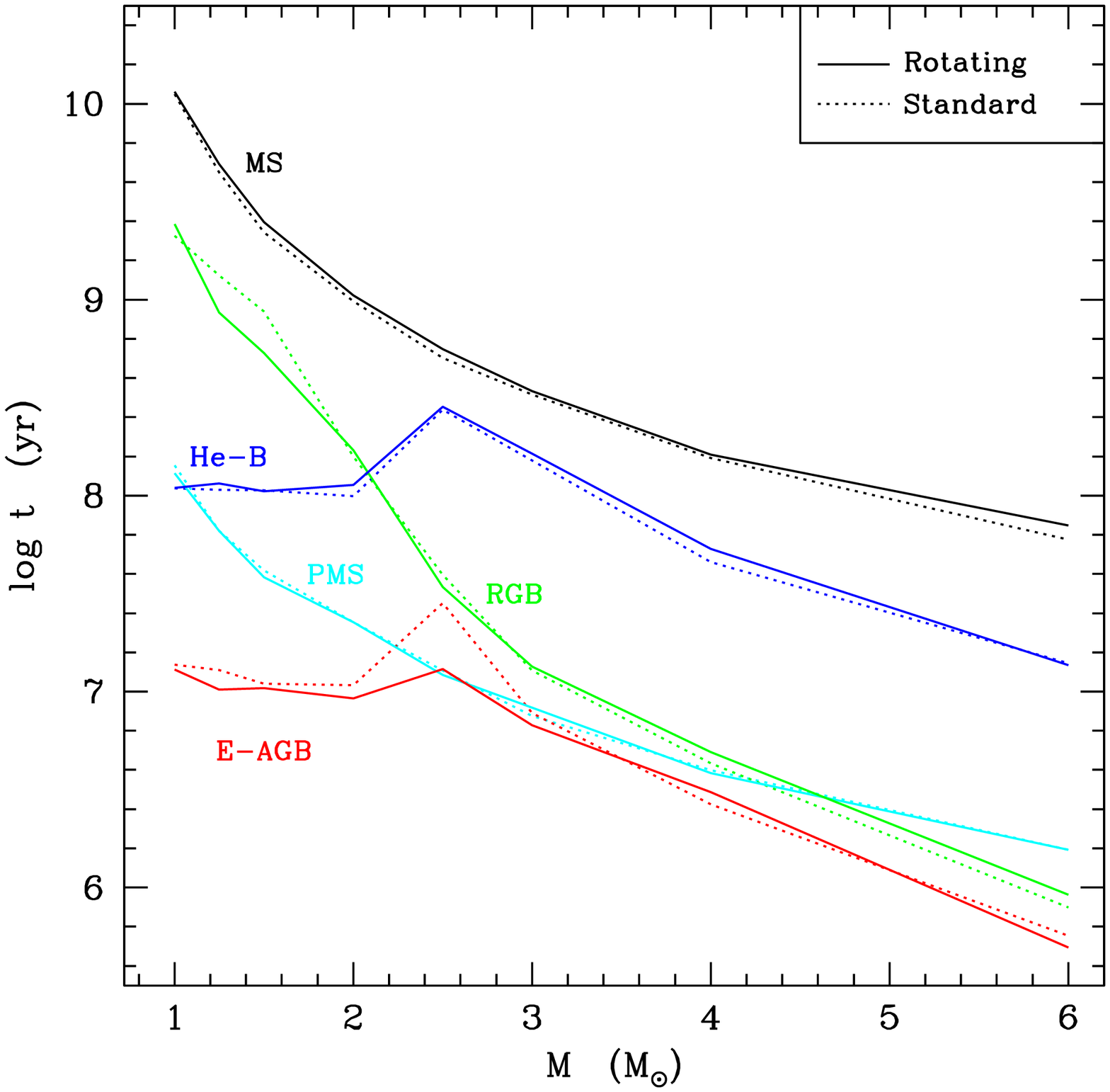}
  \caption{ Duration of the different evolution phases as a function of the initial stellar mass. 
  {\sl (left)} Effects of metallicity for the standard models. {\sl (right}) Effects of rotation at solar metallicity.}
  \label{fig:lifet}
\end{figure*}

The theoretical lifetimes are shown in Fig.\ref{fig:lifet} for the main phases of evolution as a function of the initial stellar mass and for the four metallicities considered, with the effects of rotation being presented. 
Let us recall the main points:

\begin{itemize}

\item The duration of all the evolutionary phases decreases when the initial stellar mass increases and when the metallicity decreases \citep{Schalleretal1992, Schaerer93, Charbonnel94}, which is consistent with the above-mentioned luminosity and effective-temperature differences.

\item After the bump luminosity the thermohaline mixing
      does not affect the lifetimes on the RGB and on the early-AGB phases. 

\item The lifetimes of rotating models on the main sequence are increased compared to those of the standard models. Indeed rotation-induced mixing brings fresh hydrogen fuel in the stellar core during that phase. As a consequence, the exhaustion of hydrogen in the central region is delayed and the lifetime on the main sequence increases; additionally the mass of the He-core is larger at the end of main sequence when rotation is accounted for
 \citep{Ekstrom11,HeLa00,MaMe00,MeMa00}.

\item As a consequence of having a more
      massive core at the end of the main sequence due to rotation,
      the models that undergo the He-Flash spend a shorter time on the
      RGB (see Fig.~\ref{fig:lifet}). Similarly to the MS, all
      rotating models have a longer lifetime during the
     quiescent central He burning phase. 

\item Lifetimes on the early-AGB are longer in rotating models than in standard models, due to their more massive core.  
This remains unchanged if we add the lifetime on the TP-AGB (1.1.10$^{6}$ yr for [1.5 M$_{\odot}$, Z$_{\odot}$] and 2.10$^{6}$ yr for [3 M$_{\odot}$, Z$_{\odot}$] ) to the early-AGB as the total lifetime on the AGB is only increased by a few percent.

\item As seen in previous subsections thermohaline mixing has a negligible impact on the stellar structure, and does not modify the stellar evolutionary tracks. Similarly it does not impact on lifetimes. 

\end{itemize}

\section{Global asteroseismic quantities}

In recent years, a large number of asteroseismic observations
have been obtained for different kinds of stars. In particular, the detection and characterization of solar-like oscillations in a
large number of red giants by space missions \cite[e.g.][]{DeRidder2009} promises to add valuable and independent 
constraints to current stellar models. The confrontation between models including a detailed description of
transport processes in stellar interiors and these asteroseismic constraints opens a new promising path for 
our understanding of stars. \\ 

Rotation is one of the key processes that change all outputs of stellar models (see Sect. \ref{globalppt}) with a significant impact on asteroseismic observables. 
In the case of main-sequence solar-type stars, rotation is found to shift the evolutionary tracks to the blue part of the HR diagram 
resulting in higher values of the large frequency separation for rotating models than for non-rotating ones 
at a given evolutionary stage \cite[][]{Eggenberger10a}. For red giants, rotating models are found to decrease the determined 
value of the stellar mass of a star located at a given luminosity in the HR diagram 
and to increase the value of its age. Consequently, the inclusion of rotation significantly 
changes the fundamental parameters determined for a star by performing an asteroseismic calibration \cite[][]{Eggenberger10b}. \\

\subsection{Scaling relations and asymptotic quantities provided for the present grid}

\begin{figure*}[t]
~\hfill Z=0.014 \hfill\hfill M=2.0~\Ms \hfill~\\
  \includegraphics[width=0.48\textwidth]{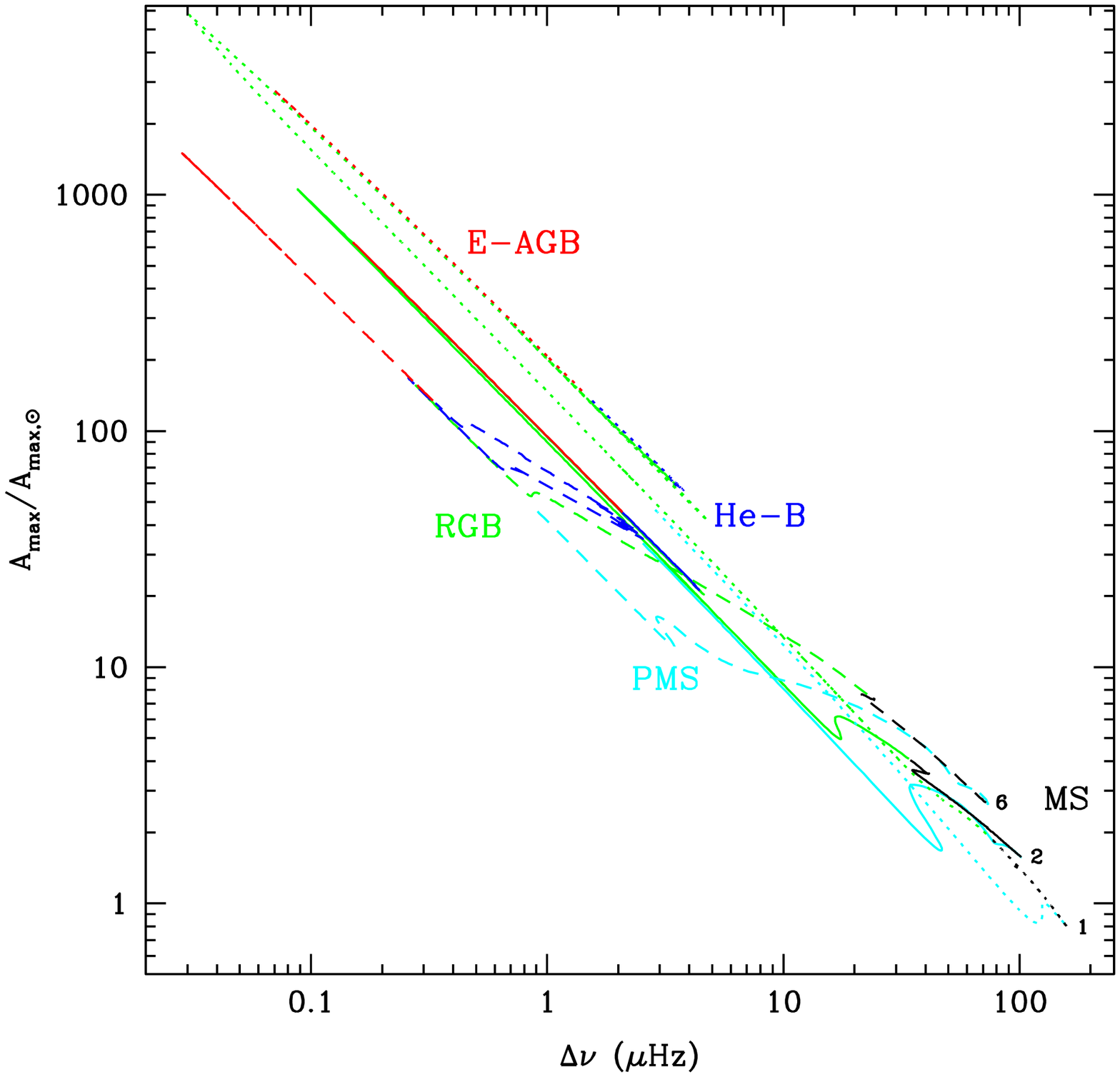}
        \includegraphics[width=0.48\textwidth]{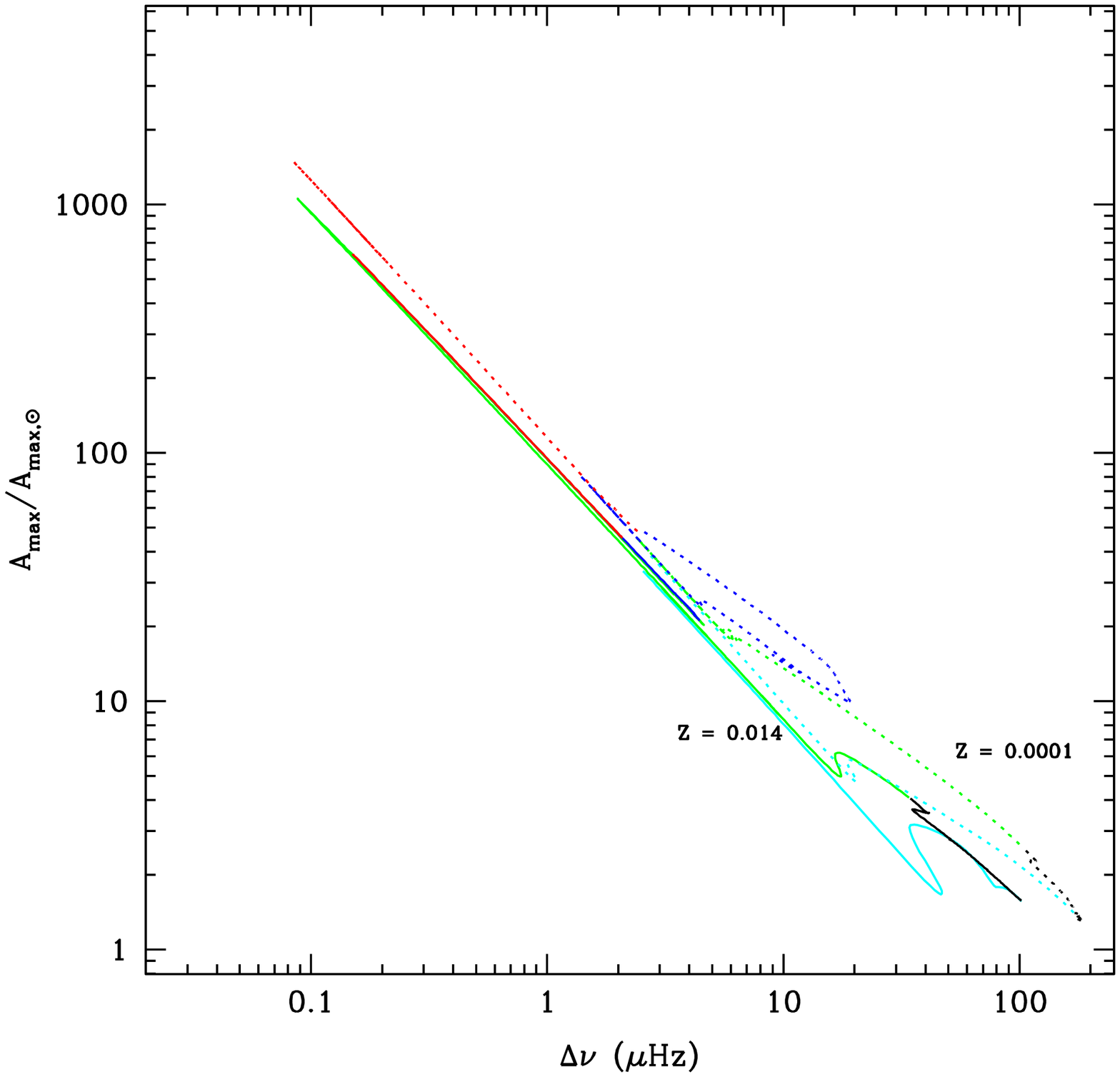}
  \caption{Maximal amplitude compared to solar as a function of the
 large separation: (\textit{left panel}) at solar metallicity for
 1.25~\Ms, 2.0~\Ms, and 6.0~\Ms ~rotating models, and (\textit{right panel }) at two metallicities Z=0.014 (solid lines) and Z=0.0001 (dotted-lines) for a 2.0~\Ms~ rotating model. Evolutionary phases are color-labeled: pre-main sequence (cyan), main sequence (black), RGB (green), He-burning (blue), and AGB (red)}
  \label{fig:asmall}
\end{figure*}

For this new grid of standard and rotating models, we now provide the values of different asteroseismic parameters (see Table \ref{tab:gridtable})
that can be directly computed from global stellar properties using
scaling relations. We also use the information on the internal
structure of models to compute asymptotic relations. These scaling relations are particularly useful 
to constrain stellar parameters and to obtain new information 
about stellar evolution without the need to perform a full asteroseismic analysis \cite[see e.g.][]{Stello2009,Miglio2009,Hekker09,Mosser10,Chaplin2011a,Beck11,Bedding11}. Although solar-like oscillations are expected to be excited in relatively cool stars (main-sequence as well as red giant stars), these global asteroseismic quantities are provided for all models of our grid and not only for models with a significant convective envelope.

The first global asteroseismic quantity we provide is the large frequency separation $\Delta \nu$, which is expected to be proportional to the square root of 
the mean stellar density \cite[e.g.][]{Ulrich1986}:

\begin{equation}
 \Delta \nu_{\mathrm{scale}} = \Delta \nu_{\odot}   \left( \frac{M}{\Ms} \right)^{0.5}  \left( \frac{R}{R_{\odot}} \right)^{-1.5},
\label{deltanuscale}
\end{equation}

with the solar large frequency separation $\Delta \nu_{\odot} = 134.9\mathrm{\mu Hz}$. \\

\citet{Belkacem2011} have proven that the frequency $\nu_{\mathrm{max}}$ at which the oscillation modes reach their strongest amplitudes is approximatively proportional to the acoustic cut-off frequency, as suspected by \citet{Brown1991} and \citet{Kjeldsen1995}. 

\begin{equation}
 \nu_{\mathrm{max}} = \nu_{\mathrm{max}, \odot}  \left( \frac{M}{\Ms}  \right)   \left(\frac{R}{R_{\odot}}   \right)^{-2}  \left(\frac{T_{\mathrm{eff}}}{T_{\mathrm{eff,\odot}}}   \right)^{-0.5},
\label{numax}
\end{equation}

with the solar value $\nu_{\mathrm{max,}\odot}=3150\mathrm{\mu Hz}$.\\

Finally, the value of maximum oscillation amplitude relative to that of the Sun ($A_{\mathrm{max}}/A_{\mathrm{max,} \odot}$) 
is computed using the following relation \cite[e.g.][]{Huberetal11} : 

\begin{equation}
 A_{\mathrm{max}}/A_{\mathrm{max,} \odot} = \frac{(L/L_{\odot})^{s}}{(M/\Ms)^{t}}  \left(  \frac{T_{\mathrm{eff}}}{T_{\mathrm{eff, \odot}}}\right)^{-r} ,
\label{Amax}
\end{equation}
with the solar value from \textit{Kepler} $A_{max, \odot}=2.5\mathrm{ppm}$.\\
Following Huber and collaborators we adopt s=0.838 and t=1.32. A value
of $r=2$ is adopted as advocated by  \cite{Kjeldsen1995}. Note
that we simply assume that these relations hold for the different models computed here and we do not test the validity of the scaling relations. Preliminary studies suggest that these simple scaling relations hold reasonnably well for main sequence and RGB stars \citep[e.g.][]{Stello2009,White2011}.\\ In addition to the asteroseismic observables deduced from scaling relations, asymptotic asteroseismic quantities are also provided. \\

The asymptotic large frequency separation is given by :

\begin{equation}
 \Delta \nu_{\mathrm{asymp}} =  \left (2 \int_0^R \frac{dr}{c_{s}} \right)^{-1},
\label{deltanuasymp}
\end{equation}
where R is the stellar radius, and c$_{s}$ is the sound speed. \\

The total acoustic radius T is provided using the following relation, which is directly related to the large frequency separation : 

\begin{equation}
T = \int_0^R \frac{dr}{c_{s}}=\frac{1}{2\cdot  \Delta \nu_{\mathrm{asymp}} },
\label{deltanu}
\end{equation}

The acoustic radius at the base of convective envelope (t$_{\mathrm{BCE}}$) and at the location of helium second-ionisation region\footnote{The Schwarzschild citerium allows us to define the base of the convective envelope. The mininum in $\Gamma_{1}$, the adiabatic exponent, corresponding to He-II defines the location of helium second ionisation region.} (t$_{\mathrm{He}}$) are determined with following relations : 

\begin{equation}
t_{\mathrm{BCE}} = \int_0^{r_{\mathrm{BCE}}} \frac{dr}{c_{s}} \hfill, \hfill   t_{\mathrm{He}}= \int_0^{r_{\mathrm{He}}} \frac{dr}{c_{s}} ,
\label{deltanu}
\end{equation}
where r$_{\mathrm{BCE}}$ and r$_{\mathrm{He}}$ represent the stellar radius at the base of convective envelope and at the location ofhelium second-ionization zone.

The period spacing of gravity modes for $\ell$=1 and different
acoustic radii can be determined with the following asymptotic relation :

\begin{equation}
 \Delta\mathrm{\Pi}(\ell=1) =  \frac{2^{\frac{1}{2}}\cdot\pi^{2}}{\int_{r_{1}}^{r_{2}} N\cdot\frac{dr}{r}},
\label{deltanu}
\end{equation}
where N is the Brunt-V\"ais\"al\"a frequency. r$_{1}$ and r$_{2}$ define the domain (in radius) where the g modes are trapped. Within this region, the mode frequency ($\omega$) must satisfy the following conditions : 

\begin{equation}
\omega^{2} < N^{2}  
\label{condi1}
\end{equation}

and

\begin{equation}
 \omega^{2} <  S_{\mathrm{l}}^{2}=\frac{l(l+1)c_{s}^{2}}{r^{2}}
\label{condi2}
\end{equation}
with S$_{l}$ the lamb frequency.

In the following discussions, we use the large separation
determined with the scaling relation $\Delta \nu_{\rm scale}$(noted
$\Delta \nu$ in the following sections). We have compared the large
separation estimates $\Delta \nu_{\mathrm{scale}}$ (Eq.~8) and $\Delta
\nu_{\mathrm{asymp}}$ (Eq.~11) to evaluate the difference between both
expressions. On the main sequence, this difference varies between 3 and 5\% for 1.0 M$_{\odot}$ model at solar metallicity.
We find that the relative error obtained when using Eq. \ref{deltanuscale} or Eq. \ref{deltanuasymp} depends on the
stellar mass (i.e. 3-8\% for [2.0 M$_{\odot}$, Z$_{\odot}$] MS model), the metallicity (i.e. 10-12\% for the [2.0 M$_{\odot}$, Z=10$^{-4}$] MS model), and the evolutionnary phase (i.e. $<$15\% on RGB, $<$10\% on He-B both for [1.0 M$_{\odot}$, Z$_{\odot}$] model). 
A forthcoming work, out of the scope of the paper, is necessary to interpret these differences and to see how they compare to \citet{White2011}

\subsection{Evolution of the asteroseismic observables}

\begin{figure*}[t]
~\hfill Pre-Main sequence\hfill\hfill Main Sequence\hfill ~ \\
  \includegraphics[width=0.5\textwidth]{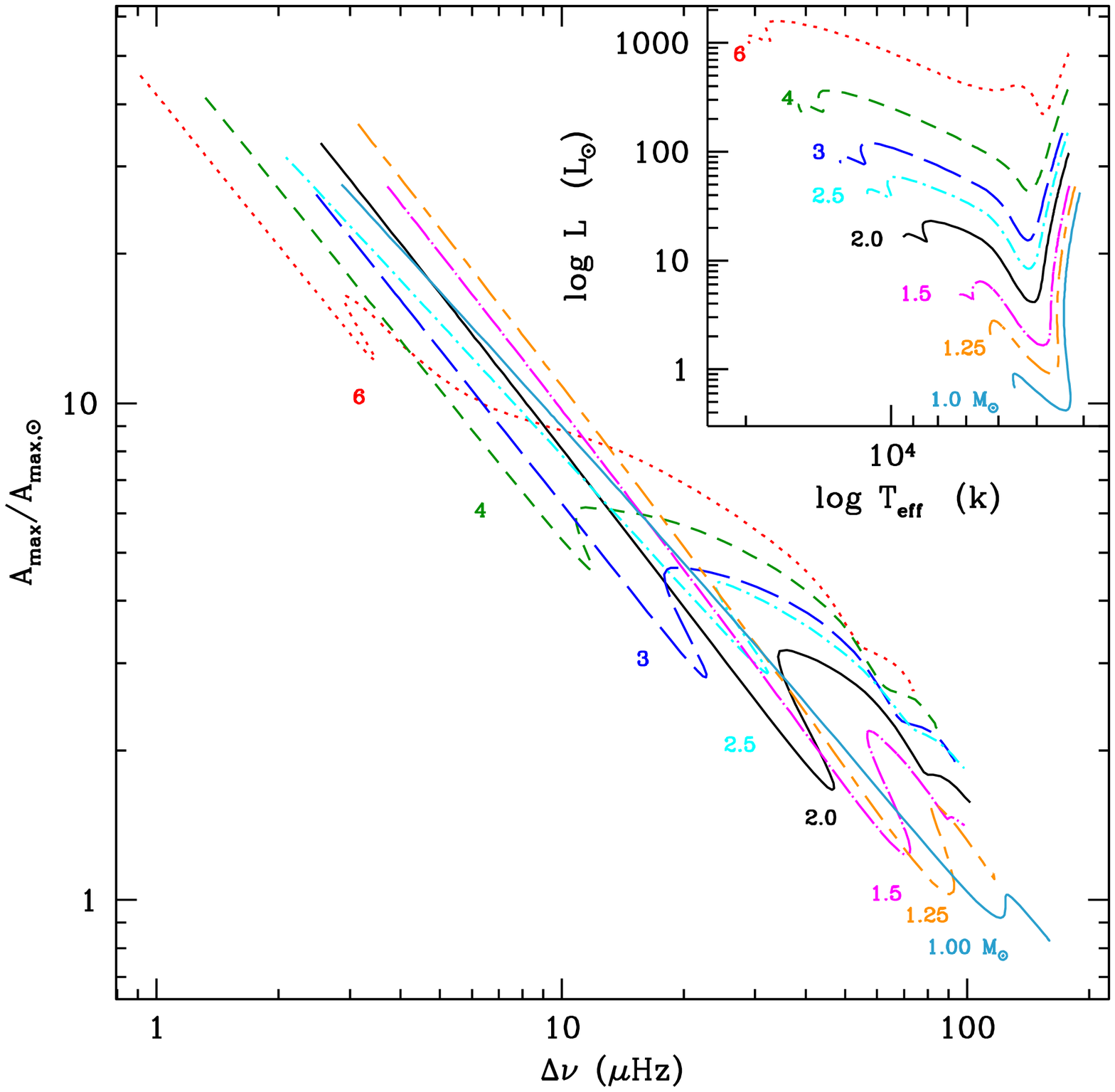}
  \includegraphics[width=0.5\textwidth]{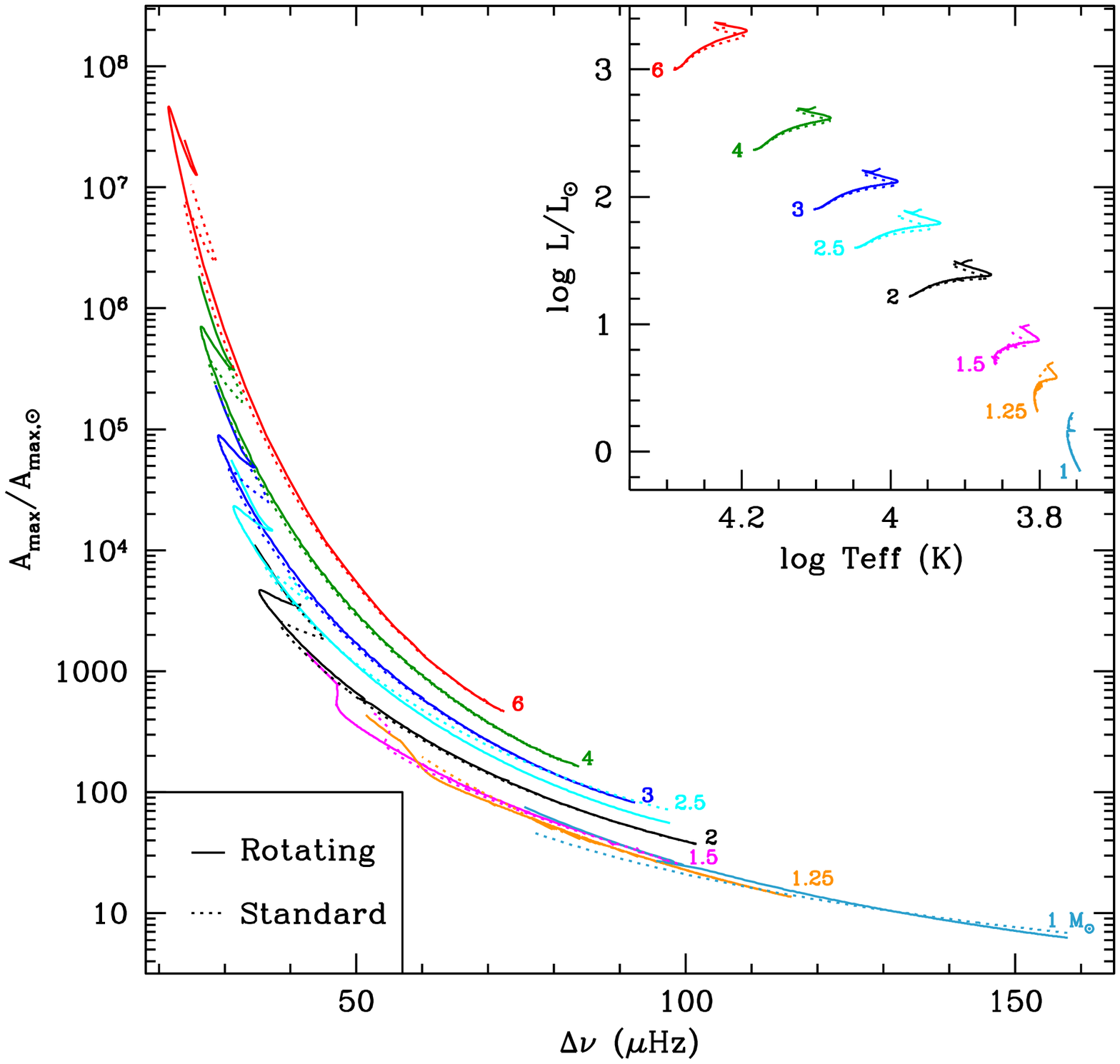}\\
 \\
~\hfill \hspace*{3.5cm} Red Giant Branch\hfill \hspace*{3.5cm}He-Burning\hfill~\\
  \includegraphics[width=0.5\textwidth]{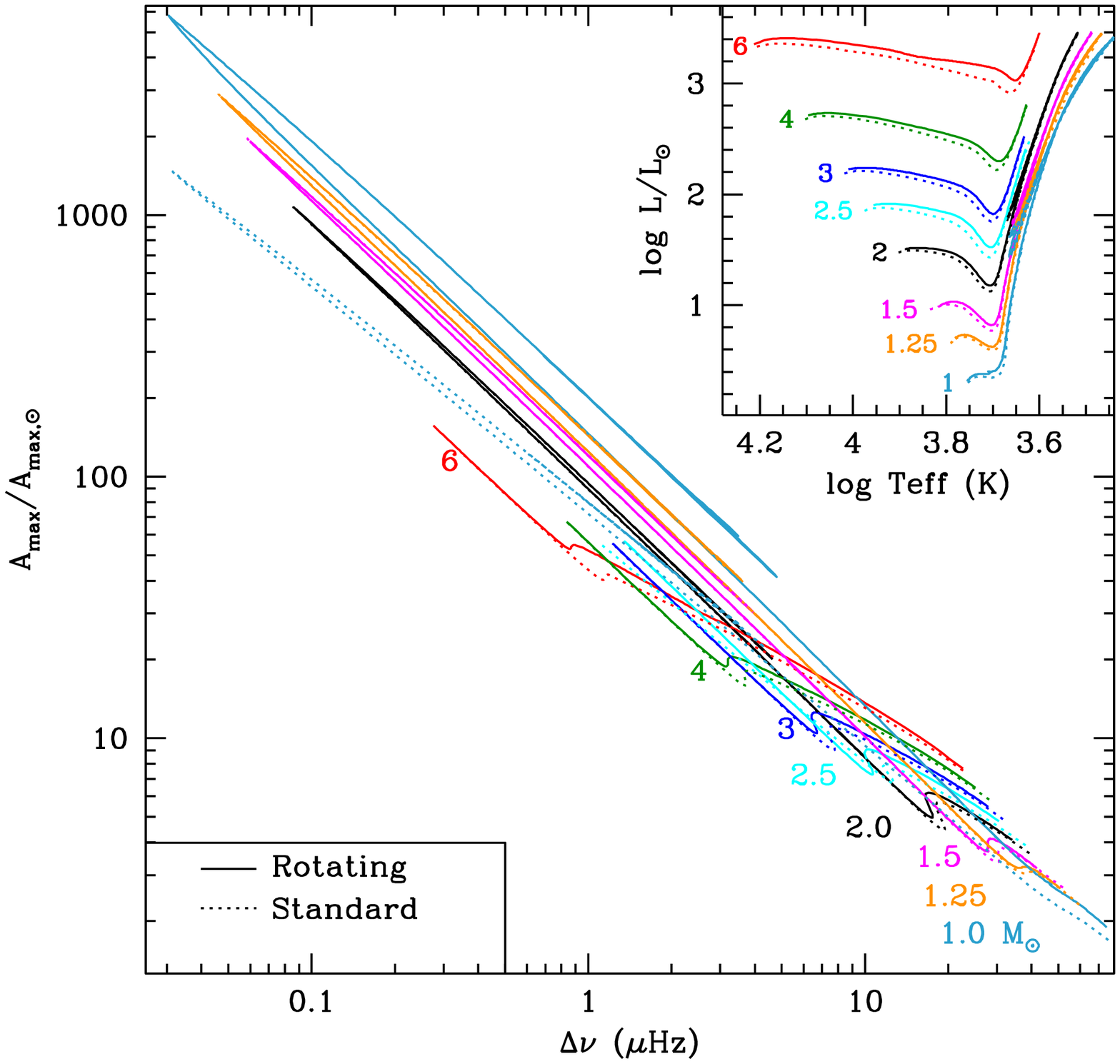}
  \includegraphics[width=0.5\textwidth]{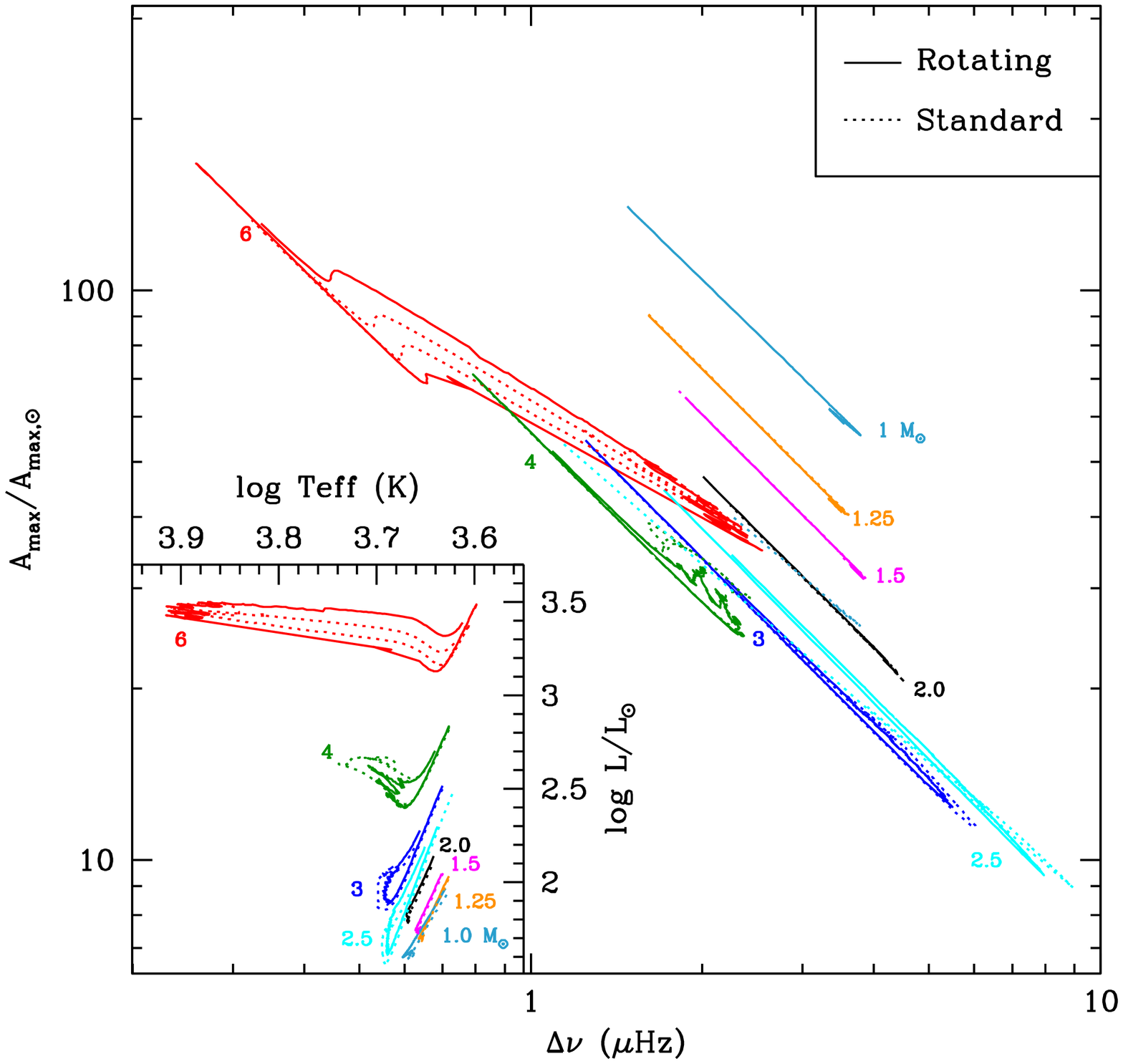}
  \caption{Maximal amplitude compared to solar as a function of the larger separation for models at \Zs. \textit{Top-left panel:} on the pre-main sequence ;  \textit{top-right panel:} on the main sequence ; \textit{bottom-left panel:} on the red giant branch ; \textit{bottom-right panel:} on He-burning phase. Standard models and rotating models are shown with dashed and solid  lines respectively except for pre-main sequence.  The corresponding evolution in the Hertzsprung-Russell diagram is shown on the right corner of each panel }
  \label{fig:asmdph}
\end{figure*}

\begin{figure}[t]
  \includegraphics[width=0.5\textwidth]{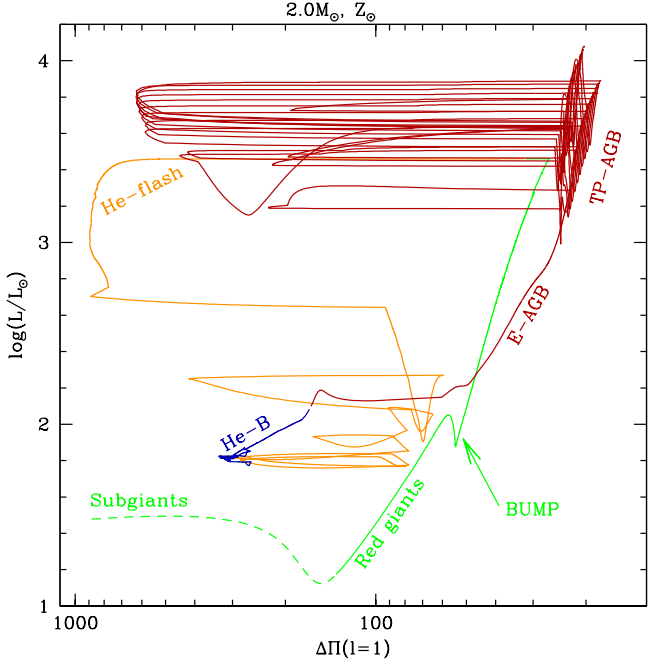}
  \caption{The stellar luminosity as a function of the asymptotic period spacing of g-modes for the standard 2.0 M$_{\odot}$ model at solar metallicity. Evolutionary phases are color-labeled: subgiant (green dashed), red giant (green solid), He-flash episod (orange), He-burning (blue), and asymptotic giant branch (red)}
  \label{fig:asmDpg}
\end{figure}

\begin{figure*}[t]
 \includegraphics[width=0.5\textwidth]{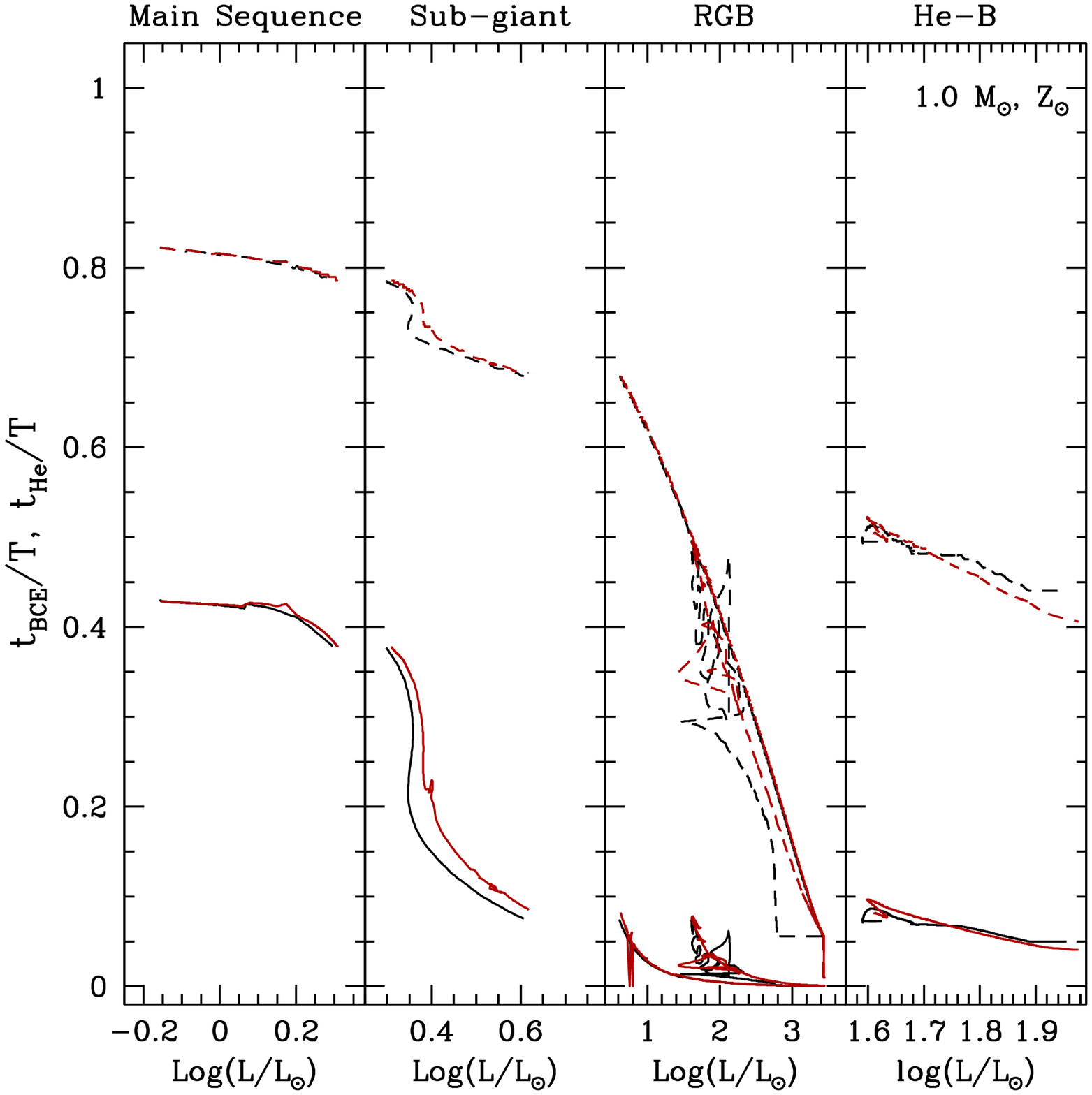}
  \includegraphics[width=0.507\textwidth, clip=true, trim= 0cm 0.3cm 0cm 0.3cm]{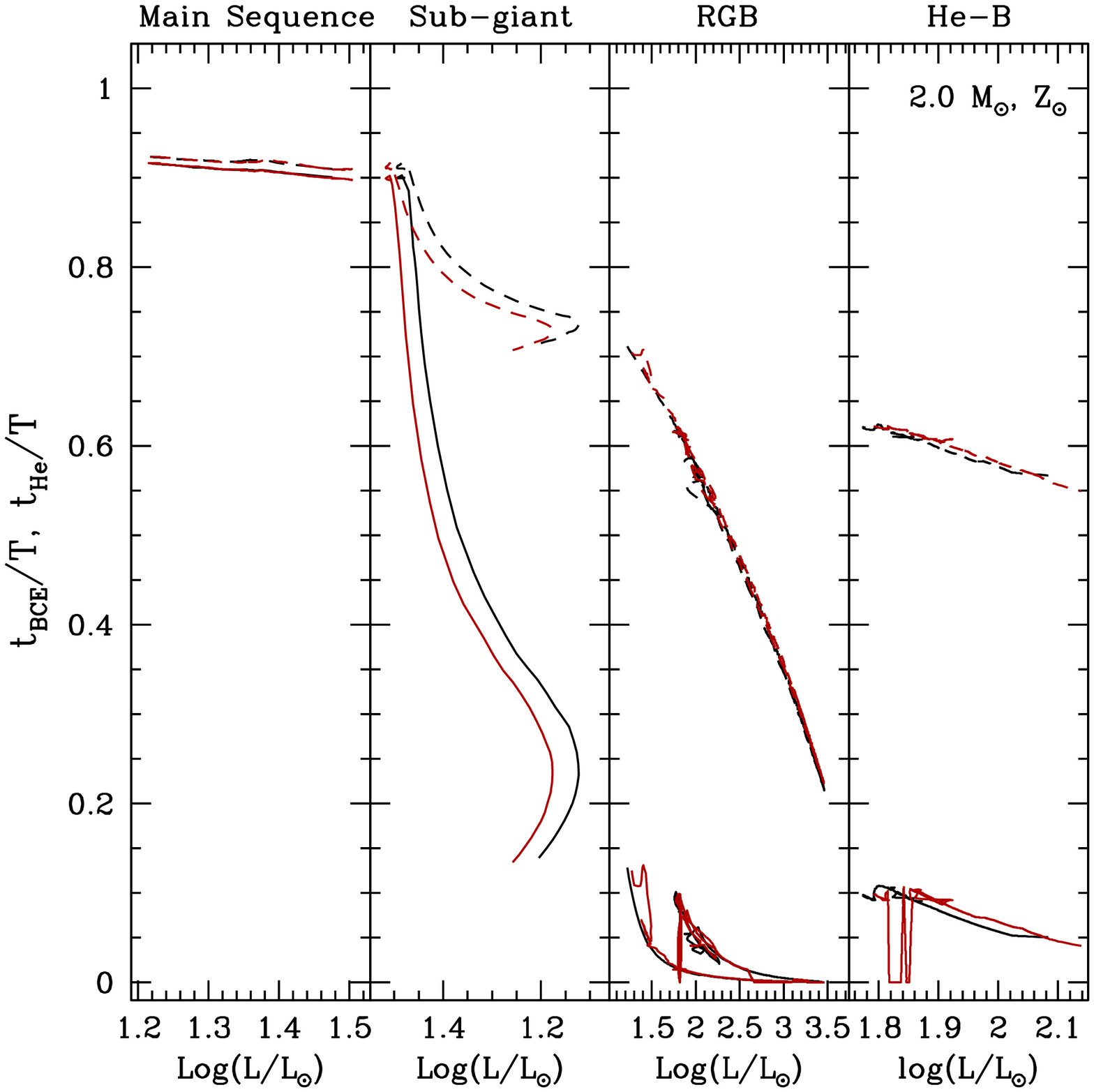}
  \caption{Acoustic radius at the base of the convective envelope (t$_{\mathrm{BCE}}$, solid line) and at the base of He-ionization zone (t$_{\mathrm{He}}$, dashed line) compared to total acoustic radius T as a function of the stellar luminosity for (left panel) 1.0 and (right panel) 2.0 M$_{\odot}$ models at solar metallicity following standard (black line) and rotating prescriptions (red line) ; (from left to right) on the main sequence,  the sub-giant branch, red giant branch, and He-burning phase.  }
  \label{fig:tbcethe}
\end{figure*}

\begin{figure*}[t]
  \includegraphics[width=0.5\textwidth]{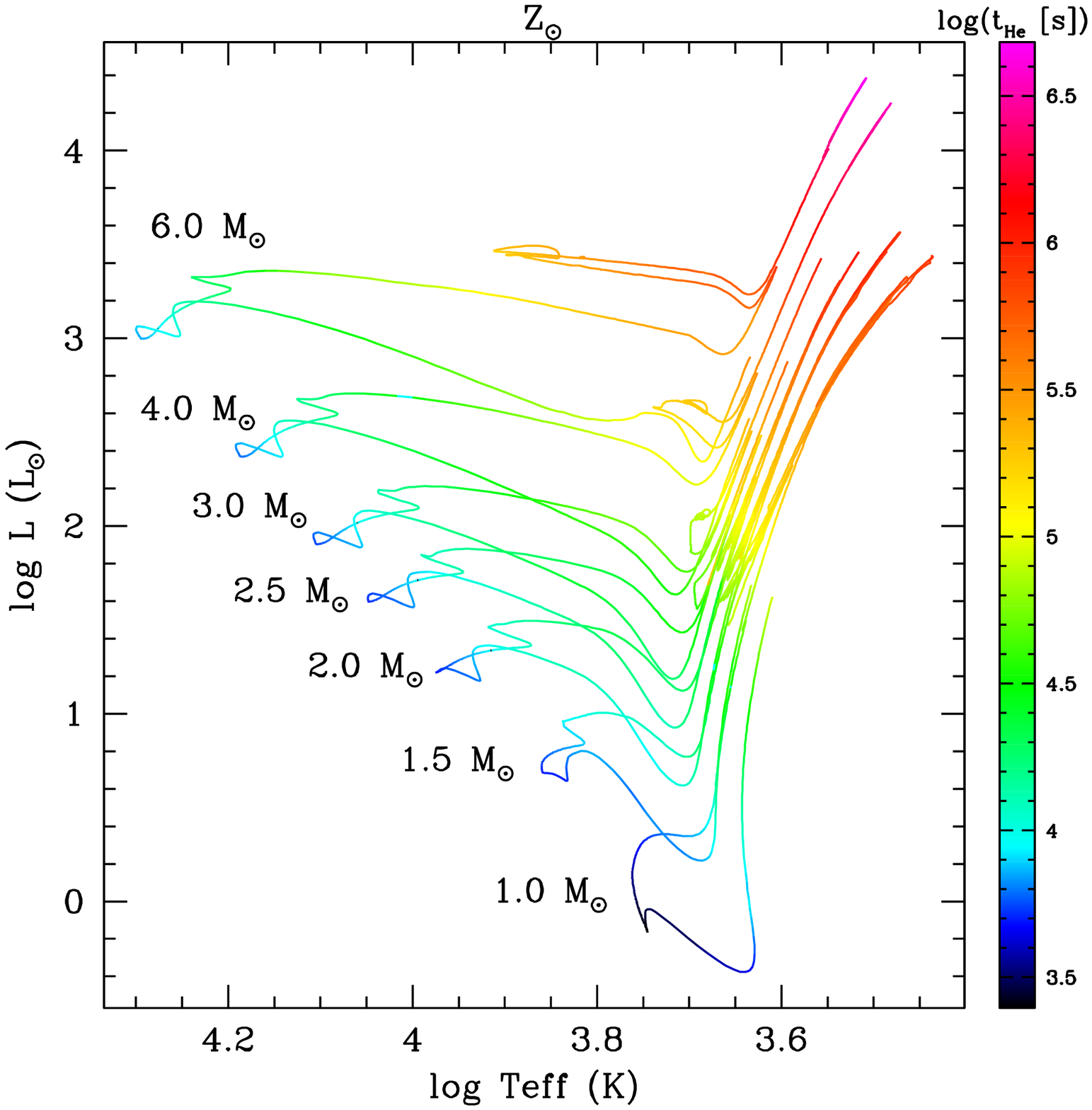}
  \includegraphics[width=0.5\textwidth]{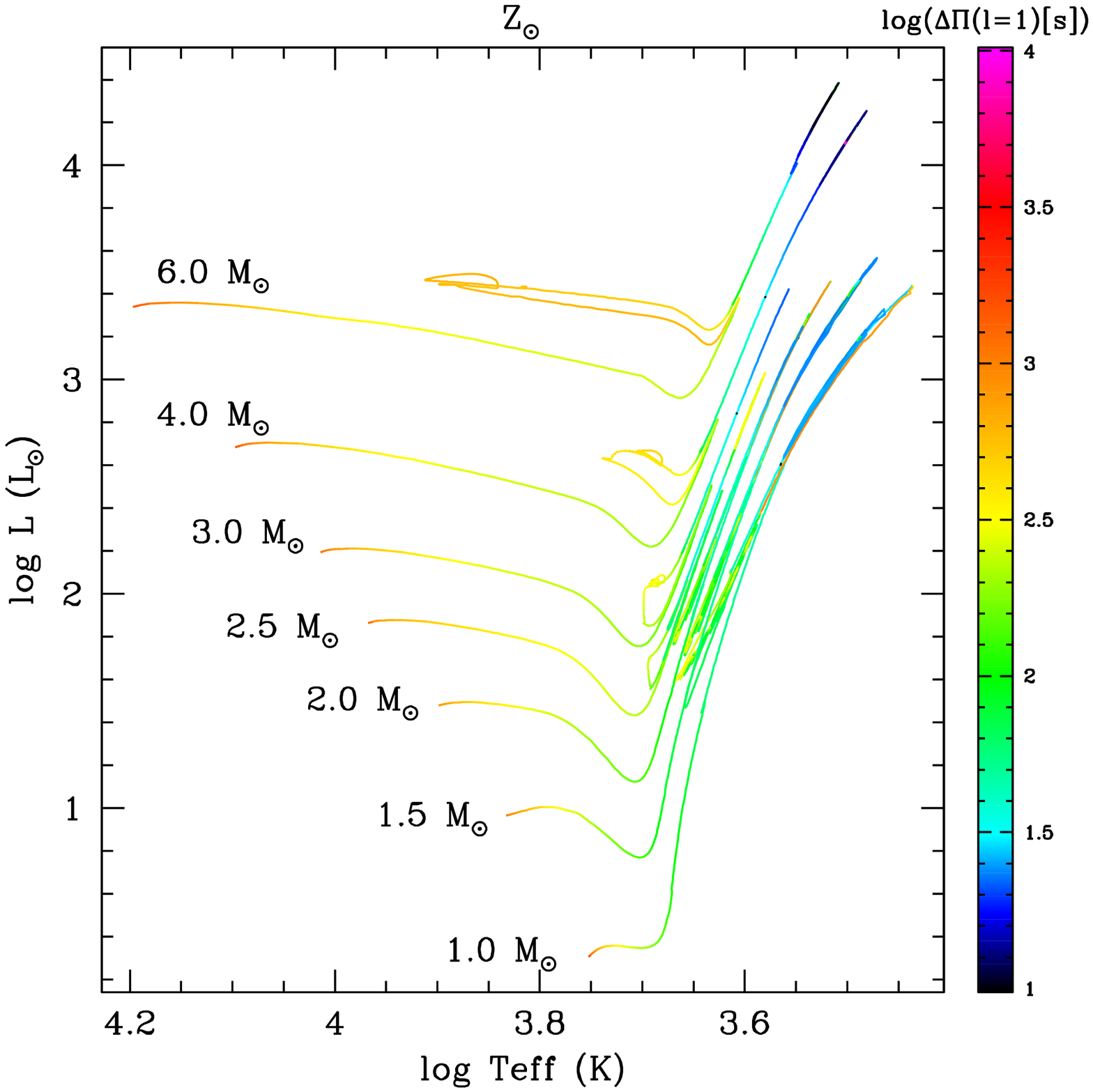}
  \caption{Color-coded HR diagram for all the masses in the
 grid at solar metallicity. The color code represents the values (in
 seconds) of the acoustic radius at the base of the ionization of He II
 $t_{\rm He}$ (left panel) and of the asymptotic period spacing of
 g-modes $\Delta\mathrm{\Pi}(\ell=1) $ (right panel). The values increase
 from blue to red in both cases as shown on the scales on the right of
 the plots.}
  \label{fig:asmHRD}
\end{figure*}

We now describe the  global asteroseismic properties of low- and intermediate-mass stars along their evolution, and discuss the impacts of stellar mass, metallicity, and of non-standard mixing mechanisms.
We illustrate our discussion with Fig. \ref{fig:asmall} to \ref{fig:asmTP}. 
In Fig.\ref{fig:asmall} we present the maximal oscillation amplitude as a function of the large frequency separation all along the evolution from the pre-main sequence to the AGB phase for three different initial stellar masses at solar metallicity (left panel) and for a 2~\Ms{} star for the lowest (Z=0.0001) and the highest (Z=0.014) metallicities explored here (right panel). The line colors change with the stellar evolution phases: cyan corresponds to the pre-main sequence, black to the main sequence, green to the red giant branch, blue to core helium-burning, and red to the asymptotic giant branch.
Fig.~\ref{fig:asmdph} presents the same quantities for all the stellar masses considered at solar metallicity,  for both the standard and the rotating cases (dashed and solid lines respectively), each evolutionary phase being presented singly.

\subsubsection{Trends for a given stellar mass and metallicity - the case of the 2.0~\Ms, ~\Zs models}

In the following we describe the evolution of the seismic properties of a star of 2.0~\Ms{} at solar metallicity as depicted by the solid line in Fig.\ref{fig:asmall} and the black line in Fig.\ref{fig:asmdph}. Although this 2\,M$_{\odot}$ star is not expected to show solar-like oscillations during the main sequence due to its too large surface temperature, it represents a typical model for a red giant exhibiting solar-like oscillations. Moreover, the discussion of the changes of the asteroseismic quantities during the main sequence remains globally valid for less massive main-sequence stars (the main difference being the value of the ratio between the acoustic radius at the base of the convective envelope and the total acoustic radius, see Fig. \ref{fig:tbcethe}).

The large separation $ \Delta \nu$ increases along the pre-main sequence because of its inverse dependence in stellar radius, which decreases 
during that phase (see Eq.\ref{deltanu}). On the other hand, the
simultaneous decrease of maximal amplitude
$A_{\mathrm{max}}/A_{\mathrm{max,}\odot}$ results from the decrease of
luminosity and from the rise of effective temperature (see Eq.\ref{Amax}).

On the main sequence, $A_{\mathrm{max}}/A_{\mathrm{max,}\odot}$ rises because of its dependence with luminosity and with the inverse of the effective temperature. In addition, the expansion of the stellar radius causes $ \Delta \nu$ to increase. 
The radius increase also causes the large separation to drop significantly as the model evolves on the RGB, while the maximum amplitude increases during this phase as a result of the luminosity increase and the effective temperature decrease.

After the He-flash at the tip of the RGB, the luminosity and the stellar
radius dwindle with a rise of effective temperature, implying that the
maximum of amplitude wanes while the large separation
increases. Throughout the helium burning phase, the luminosity and
stellar radius increase while the effective temperature and stellar mass
decrease. Consequently, a gain of $A_{\mathrm{max}}$ is obtained while the large
separation reduces. \\

Figure \ref{fig:asmDpg} presents the asymptotic period spacing
of gravity modes $\Delta\mathrm{\Pi}$($\ell$=1) for the standard model
of 2.0 M$_{\odot}$ at solar metallicity. As proposed by
\citet{Bedding11} and \citet{Mosser11}, this quantity allows to
distinguish two stars that have the same luminosity, one being at
the RGB bump and
the other one being at the clump undergoing central He burning. Indeed, at log(L/L$_{\odot}$) $\sim$
2.0, $\Delta\Pi$($\ell$=1)=55 s at the RGB bump, and
$\Delta\mathrm{\Pi}$($\ell$=1)=190 s in the clump. As the stellar structure, and more particularly the presence of the convective core affects the domain where the g-modes are trapped, $\Delta\mathrm{\Pi}$($\ell$=1) is larger in clump stars than in RGB stars \citep{Christensen11}. Large variations of $\Delta\mathrm{\Pi}$($\ell$=1) are observed during the thermal pulses and He-flash phase because of the formation of an intermediate convective zone during the He-flashes and thermal pulses. We consider that the modes are trapped in the outermost radiative zone leading to a large value of $\Delta\mathrm{\Pi}$($\ell$=1) \citep{Bildsten12}.

Figure \ref{fig:tbcethe} shows the acoustic radius at the base
of convective envelope (t$_{\mathrm{BCE}}$) and at the base of He-ionization zone
(t$_{\mathrm{He}}$), both over the total acoustic radius as a function of
luminosity, for the standard (black line) 1.0 M$_{\odot}$ (left panel) and 2.0 M$_{\odot}$ (right panel) models at solar metallicity from the main sequence to
He-burning phase. As the extent (in radius) of the convective envelope decreases with increasing stellar mass on the main sequence, the difference between t$_{\mathrm{He}}$
and t$_{\mathrm{BCE}}$ becomes smaller.

As the convective envelope deepens inside the star with the first
dredge-up, the acoustic radius t$_{\mathrm{BCE}}$ decreases during the subgiant
branch. During this phase, the acoustic radius at the base of He II ionization zone follows
the variation of effective temperature and t$_{\mathrm{He}}$ increases while T$_{\rm eff}$ decreases. In addition
the total acoustic radius increases because of its dependence with the
stellar radius that increases during this phase. Consequently,
t$_{\mathrm{He}}$/T decreases from 0.8 to 0.68 and from 0.9 to 0.7 for [1.0 M$_{\odot}$, Z$_{\odot}$] and [2.0 M$_{\odot}$, Z$_{\odot}$] models respectively . While the star ascends the RGB,
t$_{\mathrm{BCE}}$ follows the convective envelope and the acoustic radius
t$_{\mathrm{He}}$/T decreases to 0.2 for 2.0 M$_{\odot}$ model until the RGB tip is reached (at log L/L$_{\odot} \sim$
3.5). When the stellar luminosity decreases after the RGB tip the
convective envelope retracts in radius, and the effective temperature
increases while the star contracts. Therefore t$_{\mathrm{BCE}}$/T and t$_{\mathrm{He}}$/T
increase until the central temperature is sufficient to ignite
helium. At the end of He burning phase, the surface layers expand and
the convective envelope deepens while the effective temperature
decreases. During this phase t$_{\mathrm{BCE}}$/T and t$_{\mathrm{He}}$/T in 2.0 M$_{\odot}$ model decrease slowly
from $\sim$0.62 and $\sim$0.1 to $\sim$0.55 and $\sim$0.05
respectively.

\begin{figure}[t]
  \includegraphics[width=0.5\textwidth]{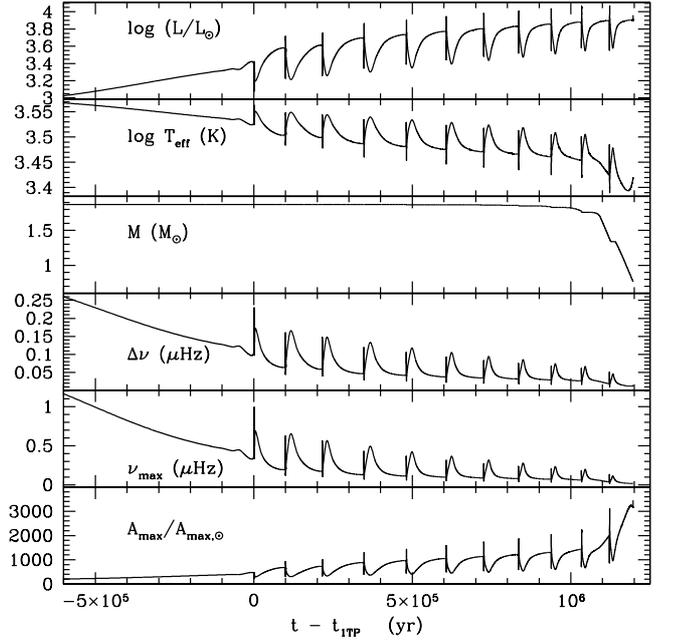}
 \caption{\textit{From top to bottom:} Evolution of global stellar
  properties (luminosity L, effective temperature T$_{\mathrm{eff}}$ and total
  stellar mass M) and asteroseismic parameters ($\Delta\nu$, and
  $\nu_{\mathrm{max}}$,A$_{\mathrm{max}}$/A$_{\mathrm{max,}\odot}$) on the TP-AGB  of the [2.0 M$_{\odot}$, Z$_{\odot}$]
  model computed with rotation up to the RGB tip and thermohaline mixing all along the evolution. The abscissa is
  the time since the first thermal pulse.
  }
  \label{fig:asmTP}
\end{figure}

Although we do not include this phase in the files presented in Sect.\ref{tablecontent}, we discuss  the evolution of the asteroseismic parameters during the TP-AGB. Figure \ref{fig:asmTP} shows the evolution of stellar parameters (\ensuremath{L},\ensuremath{T}$_{\mathrm{eff}}$, and \ensuremath{M}), and asteroseismic parameters ($\Delta\nu$, $\nu_{\mathrm{max}}$, and $A_{\mathrm{max}}$) as a function of time from the first thermal pulse for 2.0~\Ms~ model at \Zs.

Between each thermal pulse, the stellar radius increases. The mass  slowly decreases by steps at this phase, remaining almost constant during the inter-pulses.
Then, the large separation decreases between each thermal pulse as well as $\nu_{\mathrm{max}}$ despite the slight decrease of effective temperature. The maximal amplitude increases as a result of the luminosity increase and the decrease of the effective temperature and total stellar mass. The strong increase of $A_{\mathrm{max}}$ during the interpulses is due to the large mass loss at this phase, better seen in the last pulse shown in Fig. \ref{fig:asmTP}. 

\subsubsection{Effects of metallicity}
Although the current asteroseismic missions focus on solar metallicity stars, we present the effects of metallicity on the global asteroseismic parameters; they are depicted in Fig.\ref{fig:asmall} (right panel)  for a 2.0~\Ms~ model. 

 As discussed in ¤\ref{Zeffects}, at lower metallicity the main sequence is shifted to the blue and to higher luminosity on HR diagram. Therefore, the track 
 in the $A_{\mathrm{max}}$ vs $\Delta\nu$ plot is moved to lower $A_{\mathrm{max}}/A_{max,\odot}$ values. In addition, the luminosity of He-ignition at the RGB tip occurs at lower  $A_{\mathrm{max}}/A_{max,\odot}$ when the metallicity is lower. 
 The track all along the evolution in the $A_{\mathrm{max}}$ vs $\Delta\nu$ plot is shifted to the right (toward higher $\Delta \nu$ values) when the metallicity decreases. This is a direct consequence of a more metal-poor star behaving as a more massive star.
 
\subsubsection{Effects of stellar mass}

When the mass increases, the ZAMS (point at higher $\Delta\nu$ and lower $A_{\mathrm{max}}$) is shifted to lower $\Delta\nu$ and higher $A_{\mathrm{max}}$ (see Fig. \ref{fig:asmall} and top panels of Fig.\ref{fig:asmdph}). Indeed, the large frequency separation at the point of ZAMS decreases because of its dependence with the mean stellar density.
Similarly to the Hayashi tracks during the pre-main sequence, stars with different masses tend to join together in the $A_{\mathrm{max}}$ vs $\Delta\nu$ plot during the RGB and early-AGB phase (see green and red lines respectively in Fig. \ref{fig:asmall}). When stellar mass increases stars follow more extended blue loops during the central He-burning phase. As shown in Fig.~\ref{fig:asmdph} (bottom right panel), blue loops range over high $\Delta\nu$ and low $A_{\mathrm{max}}$/A$_{max,\odot}$. \\

Figure \ref{fig:asmHRD} shows the theoretical evolution of the
stellar acoustic radius at the base of He-ionization zone and of the
asymptotic period spacing of g-modes along the evolutionary tracks in
the HR diagram for models of different masses at solar metallicity.
Both quantities variations are expressed in seconds. Evolution is shown
from the pre-main sequence to the early-AGB phase and from the
subgiant branch to the early-AGB phase for t$_{\mathrm{He}}$ and
$\Delta\Pi$ respectively. At a given evolutionary phase, the larger
the stellar mass, the larger also the acoustic
radius at the base of He-ionization zone. The same is true for $\Delta\Pi$.

\subsubsection{Impact of rotation on the evolution along the amplitude vs $\Delta\nu$ diagram} \label{rotation}

In this section, we focus on the effect of accounting for rotational
mixing on the global asteroseismic parameters. As thermohaline mixing solely affects the abundance at the surface and at the external layers of HBS pattern during the giant phase, without modifying the temperature, radius, luminosity or mass of the models,  this process has no impact on the global asteroseismic parameters. It is nevertheless not excluded that well chosen asteroseismic parameters might help constraining the thermohaline mixing.

As discussed in Sect. \ref{globalppt}, rotation modifies the position of the
evolution tracks in the HR diagram (see
Fig.~\ref{fig:HRD}). Consequently, differences appear on the maximal
amplitude and the large separation, between standard and rotating
models. This effect clearly shows up at the end of the main sequence due
to the increased width of the main sequence in rotating models.  \\
Then on the subgiant branch rotating models evolve at higher
luminosities than their standard counterparts, which implies higher
maximal amplitude. Rotation has no impact on the ratios
t$_{\mathrm{BCE}}$/T and t$_{\mathrm{He}}$/T as seen in Fig. \ref{fig:tbcethe}.  In this
figure, the effect of rotation on the stellar luminosity 
during the subgiant phase is also shown.\\
When the stars reach the RGB, the differences in effective temperature and luminosity between standard and rotating models become marginal; therefore the tracks are almost identical in the Figure \ref{fig:asmdph} (bottom left panel).  \\
Rotation affects the extension and width of the blue loops. As a consequence, the  differences in asteroseismic parameters appear during the combustion phase of He, as illustrated in Fig. \ref{fig:asmdph} (bottom right panel).

\subsection{Same point on HR-diagram, same asteroseismic parameters ? }

\begin{figure}[t]
  \includegraphics[width=0.5\textwidth]{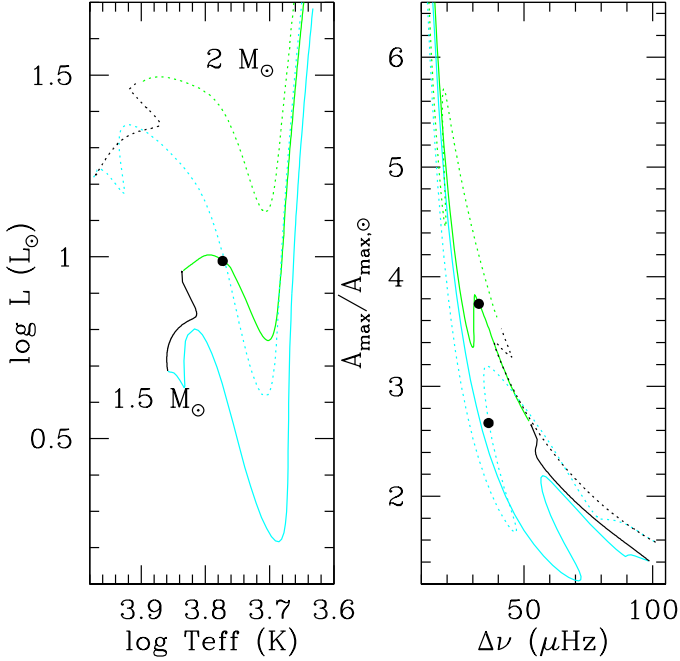}
    \caption{\textit{Left panel :} Evolution tracks in Hertzsprung-Russel diagram of 1.5 and 2.0~\Zs (solid and dotted lines respectively) stars at solar metallicity. \textit{Right panel :} Maximal amplitude over sun value as a function of the large separation  for the same models. As in figure \ref{fig:asmall}, the lines color indicates the evolutionary phase}
    \label{fig:asmAC}
\end{figure}

Figure \ref{fig:asmAC}
shows how asteroseismic parameters can help to distinguish two stars with the same effective temperature and luminosity. 
On the left panel we show the 1.5~\Ms{} and 2.0~\Ms{} tracks in the HR diagram with the intersection indicated by the black dot. At this point, the 1.5~\Ms{} star is evolving on the HR gap and has the same effective temperature and luminosity as the 2.0~\Ms~star evolving on the pre-main sequence. This black dot corresponds to two different points in the $A_{\mathrm{max}}$ vs $\Delta\nu$ plot   
shown on the right panel. 
As we can see, the larger separation of the 1.5~\Ms~ model is lower that of the 2.0~\Ms, while the maximal amplitude is higher. These stars do not have the same mean stellar density, and therefore the same $\Delta\nu$. On the other hand, the total stellar mass is different at this point, and hence the maximal amplitude.  

\section{Conclusions}

In this study we present a grid of single star evolution models in the mass range between 0.85~\Ms~ and 6.0~\Ms, for four metallicities,  including the impact of rotation and thermohaline mixing, along with standard models.
All data detailed in Table \ref{tab:gridtable} are available on the website \url{http://obswww.unige.ch/Recherche/evol/-Database-}
for  all the models computed (i.e., standard on one hand, and with rotation-induced mixing and thermohaline instability on the other hand) .\\

We recall the various impacts of metallicity variations and
rotation-induced mixing on the stellar properties that were already
discussed in literature. Thermohaline mixing does not change the stellar
parameters like luminosity and effective temperature. As such this
process does not affect the seismic properties as analyzed here. However and as
discussed in Paper I and II, it changes the surface abundances from the
bump luminosity on, and has to be taken into account to explain the
observed chemical properties of bright giant stars.  \\

Last but not least we also present the evolution of global
asteroseismic parameters for all the
models in our grid. The large
frequency separation $\Delta\nu$, the frequency $\nu_{\mathrm{max}}$, and the
maximum oscillation amplitude $A_{\mathrm{max}}$, are computed using scaling
relations. Asymptotic asteroseismic quantities are also computed : $\Delta
\nu_{\rm asympt}$, another estimate or the large frequency separation; $t_{\rm BCE}$, the acoustic radius at the base of convective envelope;
$t_{\rm He}$, the acoustic radius at the base of the HeII ionization
zone ; $T$, the total acoustic radius; and
$\Delta\mathrm{\Pi}$($\ell$=1) , the period spacing of gravity modes.

 We show that
rotation-induced mixing has an impact on these quantities, contrary to
thermohaline mixing.  While rotation changes the global properties of
main sequence stars and has an impact on the global asteroseismic
properties, thermohaline mixing is negligible on these aspects although
it changes the surface abundances in the red giants. In addition to
spectrophotometric studies, seismic studies allow to distinguish two
stars with approximatively the same luminosity and effective temperature
but with different evolutionary stages.

\begin{acknowledgements}
We wish to thank the referee and Benoit Mosser
for helpful comments on our manuscript.
We acknowledge financial support from the Swiss National Science Foundation (FNS), from ESF-Euro Genesis, and the french Programme National de Physique Stellaire (PNPS) of CNRS/INSU.
\end{acknowledgements}

\bibliographystyle{aa}
\bibliography{Reference}

\begin{thebibliography}{104}
\expandafter\ifx\csname natexlab\endcsname\relax\def\natexlab#1{#1}\fi

\bibitem[{{Aikawa} {et~al.}(2005){Aikawa}, {Arnould}, {Goriely}, {Jorissen}, \&
  {Takahashi}}]{AikawaArnould2005}
{Aikawa}, M., {Arnould}, M., {Goriely}, S., {Jorissen}, A., \& {Takahashi}, K.
  2005, \aap, 441, 1195

\bibitem[{{Arnould} {et~al.}(1999){Arnould}, {Goriely}, \&
  {Jorissen}}]{ArnouldGoriely1999}
{Arnould}, M., {Goriely}, S., \& {Jorissen}, A. 1999, \aap, 347, 572

\bibitem[{{Asplund} {et~al.}(2009){Asplund}, {Grevesse}, {Sauval}, \&
  {Scott}}]{Asplund09}
{Asplund}, M., {Grevesse}, N., {Sauval}, A.~J., \& {Scott}, P. 2009, \araa, 47,
  481

\bibitem[{{Bao} {et~al.}(2000){Bao}, {Beer}, {K{\"a}ppeler}, {Wisshak}, {Voss},
  \& {Rauscher}}]{Bao01}
{Bao}, Z.~Y., {Beer}, H., {K{\"a}ppeler}, F., {et~al.} 2000, in American
  Institute of Physics Conference Series, Vol. 529, American Institute of
  Physics Conference Series (Santa Fe, New Mexico: AIP), 706--709

\bibitem[{{Beck} {et~al.}(2011){Beck}, {Bedding}, {Mosser}, {Stello}, {Garcia},
  {Kallinger}, {Hekker}, {Elsworth}, {Frandsen}, {Carrier}, {De Ridder},
  {Aerts}, {White}, {Huber}, {Dupret}, {Montalb{\'a}n}, {Miglio}, {Noels},
  {Chaplin}, {Kjeldsen}, {Christensen-Dalsgaard}, {Gilliland}, {Brown},
  {Kawaler}, {Mathur}, \& {Jenkins}}]{Beck11}
{Beck}, P.~G., {Bedding}, T.~R., {Mosser}, B., {et~al.} 2011, Science, 332, 205

\bibitem[{{Bedding} {et~al.}(2010){Bedding}, {Huber}, {Stello}, {Elsworth},
  {Hekker}, {Kallinger}, {Mathur}, {Mosser}, {Preston}, {Ballot}, {Barban},
  {Broomhall}, {Buzasi}, {Chaplin}, {Garc{\'{\i}}a}, {Gruberbauer}, {Hale}, {De
  Ridder}, {Frandsen}, {Borucki}, {Brown}, {Christensen-Dalsgaard},
  {Gilliland}, {Jenkins}, {Kjeldsen}, {Koch}, {Belkacem}, {Bildsten}, {Bruntt},
  {Campante}, {Deheuvels}, {Derekas}, {Dupret}, {Goupil}, {Hatzes}, {Houdek},
  {Ireland}, {Jiang}, {Karoff}, {Kiss}, {Lebreton}, {Miglio}, {Montalb{\'a}n},
  {Noels}, {Roxburgh}, {Sangaralingam}, {Stevens}, {Suran}, {Tarrant}, \&
  {Weiss}}]{Bedding2010}
{Bedding}, T.~R., {Huber}, D., {Stello}, D., {et~al.} 2010, \apjl, 713, L176

\bibitem[{{Bedding} {et~al.}(2011){Bedding}, {Mosser}, {Huber},
  {Montalb{\'a}n}, {Beck}, {Christensen-Dalsgaard}, {Elsworth},
  {Garc{\'{\i}}a}, {Miglio}, {Stello}, {White}, {De Ridder}, {Hekker}, {Aerts},
  {Barban}, {Belkacem}, {Broomhall}, {Brown}, {Buzasi}, {Carrier}, {Chaplin},
  {di Mauro}, {Dupret}, {Frandsen}, {Gilliland}, {Goupil}, {Jenkins},
  {Kallinger}, {Kawaler}, {Kjeldsen}, {Mathur}, {Noels}, {Aguirre}, \&
  {Ventura}}]{Bedding11}
{Bedding}, T.~R., {Mosser}, B., {Huber}, D., {et~al.} 2011, \nat, 471, 608

\bibitem[{{Beer} {et~al.}(2002){Beer}, {Sedyshev}, {Rochow}, {Mohr}, \&
  {Oberhummer}}]{Beer02}
{Beer}, H., {Sedyshev}, P.~V., {Rochow}, W., {Mohr}, P., \& {Oberhummer}, H.
  2002, Nuclear Physics A, 705, 239

\bibitem[{{Belkacem} {et~al.}(2011){Belkacem}, {Goupil}, {Dupret}, {Samadi},
  {Baudin}, {Noels}, \& {Mosser}}]{Belkacem2011}
{Belkacem}, K., {Goupil}, M.~J., {Dupret}, M.~A., {et~al.} 2011, \aap, 530,
  A142

\bibitem[{{Bildsten} {et~al.}(2012){Bildsten}, {Paxton}, {Moore}, \&
  {Macias}}]{Bildsten12}
{Bildsten}, L., {Paxton}, B., {Moore}, K., \& {Macias}, P.~J. 2012, \apjl, 744,
  L6

\bibitem[{{Brott} {et~al.}(2011){Brott}, {de Mink}, {Cantiello}, {Langer}, {de
  Koter}, {Evans}, {Hunter}, {Trundle}, \& {Vink}}]{Brott2011}
{Brott}, I., {de Mink}, S.~E., {Cantiello}, M., {et~al.} 2011, \aap, 530, A115+

\bibitem[{{Brown} {et~al.}(1991){Brown}, {Gilliland}, {Noyes}, \&
  {Ramsey}}]{Brown1991}
{Brown}, T.~M., {Gilliland}, R.~L., {Noyes}, R.~W., \& {Ramsey}, L.~W. 1991,
  \apj, 368, 599

\bibitem[{{Brun} \& {Palacios}(2009)}]{BrunPalacios09}
{Brun}, A.~S. \& {Palacios}, A. 2009, \apj, 702, 1078

\bibitem[{{Cantiello} \& {Langer}(2010)}]{CantielloLanger2010}
{Cantiello}, M. \& {Langer}, N. 2010, \aap, 521, A9

\bibitem[{{Cassisi} {et~al.}(2006){Cassisi}, {Pietrinferni}, {Salaris},
  {Castelli}, {Cordier}, \& {Castellani}}]{Cassisi06}
{Cassisi}, S., {Pietrinferni}, A., {Salaris}, M., {et~al.} 2006, \memsai, 77,
  71

\bibitem[{{Caughlan} \& {Fowler}(1988)}]{CauFow88}
{Caughlan}, G.~R. \& {Fowler}, W.~A. 1988, Atomic Data and Nuclear Data Tables,
  40, 283

\bibitem[{{Chaboyer} \& {Zahn}(1992)}]{Chaboyer92}
{Chaboyer}, B. \& {Zahn}, J.-P. 1992, \aap, 253, 173

\bibitem[{{Chaplin} {et~al.}(2010){Chaplin}, {Appourchaux}, {Elsworth},
  {Garc{\'{\i}}a}, {Houdek}, {Karoff}, {Metcalfe}, {Molenda-{\.Z}akowicz},
  {Monteiro}, {Thompson}, {Brown}, {Christensen-Dalsgaard}, {Gilliland},
  {Kjeldsen}, {Borucki}, {Koch}, {Jenkins}, {Ballot}, {Basu}, {Bazot},
  {Bedding}, {Benomar}, {Bonanno}, {Brand{\~a}o}, {Bruntt}, {Campante},
  {Creevey}, {Di Mauro}, {Do{\u g}an}, {Dreizler}, {Eggenberger}, {Esch},
  {Fletcher}, {Frandsen}, {Gai}, {Gaulme}, {Handberg}, {Hekker}, {Howe},
  {Huber}, {Korzennik}, {Lebrun}, {Leccia}, {Martic}, {Mathur}, {Mosser},
  {New}, {Quirion}, {R{\'e}gulo}, {Roxburgh}, {Salabert}, {Schou}, {Sousa},
  {Stello}, {Verner}, {Arentoft}, {Barban}, {Belkacem}, {Benatti}, {Biazzo},
  {Boumier}, {Bradley}, {Broomhall}, {Buzasi}, {Claudi}, {Cunha}, {D'Antona},
  {Deheuvels}, {Derekas}, {Garc{\'{\i}}a Hern{\'a}ndez}, {Giampapa}, {Goupil},
  {Gruberbauer}, {Guzik}, {Hale}, {Ireland}, {Kiss}, {Kitiashvili},
  {Kolenberg}, {Korhonen}, {Kosovichev}, {Kupka}, {Lebreton}, {Leroy},
  {Ludwig}, {Mathis}, {Michel}, {Miglio}, {Montalb{\'a}n}, {Moya}, {Noels},
  {Noyes}, {Pall{\'e}}, {Piau}, {Preston}, {Roca Cort{\'e}s}, {Roth}, {Sato},
  {Schmitt}, {Serenelli}, {Silva Aguirre}, {Stevens}, {Su{\'a}rez}, {Suran},
  {Trampedach}, {Turck-Chi{\`e}ze}, {Uytterhoeven}, {Ventura}, \&
  {Wilson}}]{Chaplin2010}
{Chaplin}, W.~J., {Appourchaux}, T., {Elsworth}, Y., {et~al.} 2010, \apjl, 713,
  L169

\bibitem[{{Chaplin} {et~al.}(2011{\natexlab{a}}){Chaplin}, {Kjeldsen},
  {Bedding}, {Christensen-Dalsgaard}, {Gilliland}, {Kawaler}, {Appourchaux},
  {Elsworth}, {Garc{\'{\i}}a}, {Houdek}, {Karoff}, {Metcalfe},
  {Molenda-{\.Z}akowicz}, {Monteiro}, {Thompson}, {Verner}, {Batalha},
  {Borucki}, {Brown}, {Bryson}, {Christiansen}, {Clarke}, {Jenkins}, {Klaus},
  {Koch}, {An}, {Ballot}, {Basu}, {Benomar}, {Bonanno}, {Broomhall},
  {Campante}, {Corsaro}, {Creevey}, {Esch}, {Gai}, {Gaulme}, {Hale},
  {Handberg}, {Hekker}, {Huber}, {Mathur}, {Mosser}, {New}, {Pinsonneault},
  {Pricopi}, {Quirion}, {R{\'e}gulo}, {Roxburgh}, {Salabert}, {Stello}, \&
  {Suran}}]{Chaplin2011b}
{Chaplin}, W.~J., {Kjeldsen}, H., {Bedding}, T.~R., {et~al.}
  2011{\natexlab{a}}, \apj, 732, 54

\bibitem[{{Chaplin} {et~al.}(2011{\natexlab{b}}){Chaplin}, {Kjeldsen},
  {Christensen-Dalsgaard}, {Basu}, {Miglio}, {Appourchaux}, {Bedding},
  {Elsworth}, {Garc{\'{\i}}a}, {Gilliland}, {Girardi}, {Houdek}, {Karoff},
  {Kawaler}, {Metcalfe}, {Molenda-{\.Z}akowicz}, {Monteiro}, {Thompson},
  {Verner}, {Ballot}, {Bonanno}, {Brand{\~a}o}, {Broomhall}, {Bruntt},
  {Campante}, {Corsaro}, {Creevey}, {Do{\u g}an}, {Esch}, {Gai}, {Gaulme},
  {Hale}, {Handberg}, {Hekker}, {Huber}, {Jim{\'e}nez}, {Mathur}, {Mazumdar},
  {Mosser}, {New}, {Pinsonneault}, {Pricopi}, {Quirion}, {R{\'e}gulo},
  {Salabert}, {Serenelli}, {Aguirre}, {Sousa}, {Stello}, {Stevens}, {Suran},
  {Uytterhoeven}, {White}, {Borucki}, {Brown}, {Jenkins}, {Kinemuchi}, {Van
  Cleve}, \& {Klaus}}]{Chaplin2011a}
{Chaplin}, W.~J., {Kjeldsen}, H., {Christensen-Dalsgaard}, J., {et~al.}
  2011{\natexlab{b}}, Science, 332, 213

\bibitem[{{Charbonnel}(1994)}]{Charbonnel94}
{Charbonnel}, C. 1994, \aap, 282, 811

\bibitem[{{Charbonnel} \& {Lagarde}(2010)}]{ChaLag10}
{Charbonnel}, C. \& {Lagarde}, N. 2010, \aap, 522, A10, Paper I

\bibitem[{{Charbonnel} \& {Talon}(1999)}]{ChaTal99}
{Charbonnel}, C. \& {Talon}, S. 1999, \aap, 351, 635

\bibitem[{{Charbonnel} \& {Talon}(2005)}]{ChTa05}
{Charbonnel}, C. \& {Talon}, S. 2005, Science, 309, 2189

\bibitem[{{Charbonnel} \& {Talon}(2008)}]{ChTa08}
{Charbonnel}, C. \& {Talon}, S. 2008, in IAU Symposium, Vol. 252, IAU
  Symposium, ed. {L.~Deng \& K.~L.~Chan} (Sanya: CUP), 163--174

\bibitem[{{Charbonnel} \& {Zahn}(2007{\natexlab{a}})}]{ChaZah07b}
{Charbonnel}, C. \& {Zahn}, J. 2007{\natexlab{a}}, \aap, 476, L29

\bibitem[{{Charbonnel} \& {Zahn}(2007{\natexlab{b}})}]{ChaZah07a}
{Charbonnel}, C. \& {Zahn}, J.-P. 2007{\natexlab{b}}, \aap, 467, L15

\bibitem[{{Christensen-Dalsgaard}(2011)}]{Christensen11}
{Christensen-Dalsgaard}, J. 2011, ArXiv e-prints 1106.5946

\bibitem[{{Coc} {et~al.}(2004){Coc}, {Vangioni-Flam}, {Descouvemont},
  {Adahchour}, \& {Angulo}}]{Coc04}
{Coc}, A., {Vangioni-Flam}, E., {Descouvemont}, P., {Adahchour}, A., \&
  {Angulo}, C. 2004, \apj, 600, 544

\bibitem[{{Cunha} {et~al.}(2006){Cunha}, {Hubeny}, \& {Lanz}}]{Cunha06}
{Cunha}, K., {Hubeny}, I., \& {Lanz}, T. 2006, \apjl, 647, L143

\bibitem[{{De Ridder} {et~al.}(2009){De Ridder}, {Barban}, {Baudin}, {Carrier},
  {Hatzes}, {Hekker}, {Kallinger}, {Weiss}, {Baglin}, {Auvergne}, {Samadi},
  {Barge}, \& {Deleuil}}]{DeRidder2009}
{De Ridder}, J., {Barban}, C., {Baudin}, F., {et~al.} 2009, \nat, 459, 398

\bibitem[{{Decressin} {et~al.}(2009){Decressin}, {Mathis}, {Palacios}, {Siess},
  {Talon}, {Charbonnel}, \& {Zahn}}]{Decressin09}
{Decressin}, T., {Mathis}, S., {Palacios}, A., {et~al.} 2009, \aap, 495, 271

\bibitem[{{Denissenkov}(2010)}]{Denissenkov10}
{Denissenkov}, P.~A. 2010, \apj, 723, 563

\bibitem[{{Denissenkov} \& {Merryfield}(2011)}]{DenissenkovMerryfield10}
{Denissenkov}, P.~A. \& {Merryfield}, W.~J. 2011, \apjl, 727, L8

\bibitem[{{Descouvemont} {et~al.}(2004){Descouvemont}, {Adahchour}, {Angulo},
  {Coc}, \& {Vangioni-Flam}}]{Descouvemont04}
{Descouvemont}, P., {Adahchour}, A., {Angulo}, C., {Coc}, A., \&
  {Vangioni-Flam}, E. 2004, Atomic Data and Nuclear Data Tables, 88, 203

\bibitem[{{Dufour}(1999)}]{Dufour1999}
{Dufour}, E. 1999, PhD thesis, Universit\'e Joseph Fourier -- Grenoble I

\bibitem[{{Eggenberger} {et~al.}(2005){Eggenberger}, {Maeder}, \&
  {Meynet}}]{Eggenbergeretal05}
{Eggenberger}, P., {Maeder}, A., \& {Meynet}, G. 2005, \aap, 440, L9

\bibitem[{{Eggenberger} {et~al.}(2010{\natexlab{a}}){Eggenberger}, {Meynet},
  {Maeder}, {Miglio}, {Montalban}, {Carrier}, {Mathis}, {Charbonnel}, \&
  {Talon}}]{Eggenberger10a}
{Eggenberger}, P., {Meynet}, G., {Maeder}, A., {et~al.} 2010{\natexlab{a}},
  \aap, 519, A116

\bibitem[{{Eggenberger} {et~al.}(2010{\natexlab{b}}){Eggenberger}, {Miglio},
  {Montalban}, {Moreira}, {Noels}, {Meynet}, \& {Maeder}}]{Eggenberger10b}
{Eggenberger}, P., {Miglio}, A., {Montalban}, J., {et~al.} 2010{\natexlab{b}},
  \aap, 509, A72

\bibitem[{{Eggleton} {et~al.}(2006){Eggleton}, {Dearborn}, \&
  {Lattanzio}}]{Eggleton06}
{Eggleton}, P.~P., {Dearborn}, D.~S.~P., \& {Lattanzio}, J.~C. 2006, Science,
  314, 1580

\bibitem[{{Eggleton} {et~al.}(2008){Eggleton}, {Dearborn}, \&
  {Lattanzio}}]{Eggleton08}
{Eggleton}, P.~P., {Dearborn}, D.~S.~P., \& {Lattanzio}, J.~C. 2008, \apj, 677,
  581

\bibitem[{{Eggleton} {et~al.}(1973){Eggleton}, {Faulkner}, \&
  {Flannery}}]{EggletonFaulkner1973}
{Eggleton}, P.~P., {Faulkner}, J., \& {Flannery}, B.~P. 1973, \aap, 23, 325

\bibitem[{{Ekstr{\"o}m} {et~al.}(2012){Ekstr{\"o}m}, {Georgy}, {Eggenberger},
  {Meynet}, {Mowlavi}, {Wyttenbach}, {Granada}, {Decressin}, {Hirschi},
  {Frischknecht}, {Charbonnel}, \& {Maeder}}]{Ekstrom11}
{Ekstr{\"o}m}, S., {Georgy}, C., {Eggenberger}, P., {et~al.} 2012, \aap, 537,
  A146

\bibitem[{{Ferguson} {et~al.}(2005){Ferguson}, {Alexander}, {Allard}, {Barman},
  {Bodnarik}, {Hauschildt}, {Heffner-Wong}, \& {Tamanai}}]{Ferguson05}
{Ferguson}, J.~W., {Alexander}, D.~R., {Allard}, F., {et~al.} 2005, \apj, 623,
  585

\bibitem[{{Forestini} \& {Charbonnel}(1997)}]{FoCh97}
{Forestini}, M. \& {Charbonnel}, C. 1997, \aaps, 123, 241

\bibitem[{{Funck} \& {Langanke}(1989)}]{Funk89}
{Funck}, C. \& {Langanke}, K. 1989, \apj, 344, 46

\bibitem[{{Fynbo} {et~al.}(2005){Fynbo}, {Diget}, {Bergmann}, {Borge},
  {Cederk{\"a}ll}, {Dendooven}, {Fraile}, {Franchoo}, {Fedosseev}, {Fulton},
  {Huang}, {Huikari}, {Jeppesen}, {Jokinen}, {Jones}, {Jonson}, {K{\"o}ster},
  {Langanke}, {Meister}, {Nilsson}, {Nyman}, {Prezado}, {Riisager},
  {Rinta-Antila}, {Tengblad}, {Turrion}, {Wang}, {Weissman}, {Wilhelmsen},
  {{\"A}yst{\"o}}, \& {ISOLDE Collaboration}}]{Fynbo05}
{Fynbo}, H.~O.~U., {Diget}, C.~A., {Bergmann}, U.~C., {et~al.} 2005, \nat, 433,
  136

\bibitem[{{Graboske} {et~al.}(1973){Graboske}, {Dewitt}, {Grossman}, \&
  {Cooper}}]{Graboske73}
{Graboske}, H.~C., {Dewitt}, H.~E., {Grossman}, A.~S., \& {Cooper}, M.~S. 1973,
  \apj, 181, 457

\bibitem[{{Hale} {et~al.}(2002){Hale}, {Champagne}, {Iliadis}, {Hansper},
  {Powell}, \& {Blackmon}}]{Haleetal02}
{Hale}, S.~E., {Champagne}, A.~E., {Iliadis}, C., {et~al.} 2002, \prc, 65,
  015801

\bibitem[{{Hauser} \& {Feshbach}(1952)}]{HaFe52}
{Hauser}, W. \& {Feshbach}, H. 1952, Physical Review, 87, 366

\bibitem[{{Heger} \& {Langer}(2000)}]{HeLa00}
{Heger}, A. \& {Langer}, N. 2000, \apj, 544, 1016

\bibitem[{{Hekker} {et~al.}(2009){Hekker}, {Kallinger}, {Baudin}, {De Ridder},
  {Barban}, {Carrier}, {Hatzes}, {Weiss}, \& {Baglin}}]{Hekker09}
{Hekker}, S., {Kallinger}, T., {Baudin}, F., {et~al.} 2009, \aap, 506, 465

\bibitem[{{Horiguchi} {et~al.}(1996){Horiguchi}, {Tachibana}, \&
  {Katakura}}]{Horiguchi96}
{Horiguchi}, T., {Tachibana}, T., \& {Katakura}, J. 1996, Nuclear Data Center,
  Japan Atomic Energy Research Institute, Ibaraki

\bibitem[{{Huber} {et~al.}(2011){Huber}, {Bedding}, {Stello}, {Hekker},
  {Mathur}, {Mosser}, {Verner}, {Bonanno}, {Buzasi}, {Campante}, {Elsworth},
  {Hale}, {Kallinger}, {Silva Aguirre}, {Chaplin}, {De Ridder},
  {Garc{\'{\i}}a}, {Appourchaux}, {Frandsen}, {Houdek}, {Molenda-{\.Z}akowicz},
  {Monteiro}, {Christensen-Dalsgaard}, {Gilliland}, {Kawaler}, {Kjeldsen},
  {Broomhall}, {Corsaro}, {Salabert}, {Sanderfer}, {Seader}, \&
  {Smith}}]{Huberetal11}
{Huber}, D., {Bedding}, T.~R., {Stello}, D., {et~al.} 2011, \apj, 743, 143

\bibitem[{{Iglesias} \& {Rogers}(1996)}]{IglRog96}
{Iglesias}, C.~A. \& {Rogers}, F.~J. 1996, \apj, 464, 943

\bibitem[{{Iliadis} {et~al.}(2001){Iliadis}, {D'Auria}, {Starrfield},
  {Thompson}, \& {Wiescher}}]{Illiadisetal01}
{Iliadis}, C., {D'Auria}, J.~M., {Starrfield}, S., {Thompson}, W.~J., \&
  {Wiescher}, M. 2001, \apjs, 134, 151

\bibitem[{{Kawaler}(1988)}]{Kawaler88}
{Kawaler}, S.~D. 1988, \apj, 333, 236

\bibitem[{{Kjeldsen} \& {Bedding}(1995)}]{Kjeldsen1995}
{Kjeldsen}, H. \& {Bedding}, T.~R. 1995, \aap, 293, 87

\bibitem[{{Koehler} \& {O'brien}(1989{\natexlab{a}})}]{KoeObr89}
{Koehler}, P.~E. \& {O'brien}, H.~A. 1989{\natexlab{a}}, \prc, 39, 1655

\bibitem[{{Koehler} \& {O'brien}(1989{\natexlab{b}})}]{Koehler89}
{Koehler}, P.~E. \& {O'brien}, H.~A. 1989{\natexlab{b}}, \prc, 39, 1655

\bibitem[{{Krishnamurti}(2003)}]{Krish03}
{Krishnamurti}, R. 2003, Journal of Fluid Mechanics, 483, 287

\bibitem[{{Lagarde} {et~al.}(2011){Lagarde}, {Charbonnel}, {Decressin}, \&
  {Hagelberg}}]{Lagarde11}
{Lagarde}, N., {Charbonnel}, C., {Decressin}, T., \& {Hagelberg}, J. 2011,
  \aap, 536, A28

\bibitem[{{Maeder}(1997)}]{Maeder97}
{Maeder}, A. 1997, \aap, 321, 134

\bibitem[{{Maeder}(2009)}]{Maeder09}
{Maeder}, A. 2009, {Physics, Formation and Evolution of Rotating Stars}
  (Springer Berlin Heidelberg)

\bibitem[{{Maeder} \& {Meynet}(2000)}]{MaMe00}
{Maeder}, A. \& {Meynet}, G. 2000, \araa, 38, 143

\bibitem[{{Maeder} \& {Zahn}(1998)}]{MaeZah98}
{Maeder}, A. \& {Zahn}, J.-P. 1998, \aap, 334, 1000

\bibitem[{{Mathis} \& {Zahn}(2004)}]{MaZa04}
{Mathis}, S. \& {Zahn}, J.-P. 2004, \aap, 425, 229

\bibitem[{{Meynet} \& {Maeder}(2000)}]{MeMa00}
{Meynet}, G. \& {Maeder}, A. 2000, \aap, 361, 101

\bibitem[{{Michel} {et~al.}(2008){Michel}, {Baglin}, {Auvergne}, {Catala},
  {Samadi}, {Baudin}, {Appourchaux}, {Barban}, {Weiss}, {Berthomieu},
  {Boumier}, {Dupret}, {Garcia}, {Fridlund}, {Garrido}, {Goupil}, {Kjeldsen},
  {Lebreton}, {Mosser}, {Grotsch-Noels}, {Janot-Pacheco}, {Provost},
  {Roxburgh}, {Thoul}, {Toutain}, {Tiph{\`e}ne}, {Turck-Chieze}, {Vauclair},
  {Vauclair}, {Aerts}, {Alecian}, {Ballot}, {Charpinet}, {Hubert},
  {Ligni{\`e}res}, {Mathias}, {Monteiro}, {Neiner}, {Poretti}, {Renan de
  Medeiros}, {Ribas}, {Rieutord}, {Cort{\'e}s}, \& {Zwintz}}]{Michel08}
{Michel}, E., {Baglin}, A., {Auvergne}, M., {et~al.} 2008, Science, 322, 558

\bibitem[{{Miglio} {et~al.}(2009){Miglio}, {Montalb{\'a}n}, {Baudin},
  {Eggenberger}, {Noels}, {Hekker}, {De Ridder}, {Weiss}, \&
  {Baglin}}]{Miglio2009}
{Miglio}, A., {Montalb{\'a}n}, J., {Baudin}, F., {et~al.} 2009, \aap, 503, L21

\bibitem[{{Mitler}(1977)}]{Mitler77}
{Mitler}, H.~E. 1977, \apj, 212, 513

\bibitem[{{Mosser} {et~al.}(2011){Mosser}, {Barban}, {Montalb{\'a}n}, {Beck},
  {Miglio}, {Belkacem}, {Goupil}, {Hekker}, {De Ridder}, {Dupret}, {Elsworth},
  {Noels}, {Baudin}, {Michel}, {Samadi}, {Auvergne}, {Baglin}, \&
  {Catala}}]{Mosser11}
{Mosser}, B., {Barban}, C., {Montalb{\'a}n}, J., {et~al.} 2011, \aap, 532, A86

\bibitem[{{Mosser} {et~al.}(2010){Mosser}, {Belkacem}, {Goupil}, {Miglio},
  {Morel}, {Barban}, {Baudin}, {Hekker}, {Samadi}, {De Ridder}, {Weiss},
  {Auvergne}, \& {Baglin}}]{Mosser10}
{Mosser}, B., {Belkacem}, K., {Goupil}, M.-J., {et~al.} 2010, \aap, 517, A22

\bibitem[{{Mukhamedzhanov} {et~al.}(2003){Mukhamedzhanov}, {B{\'e}m}, {Brown},
  {Burjan}, {Gagliardi}, {Kroha}, {Nov{\'a}k}, {Nunes}, {Isko{\v r}},
  {Pirlepesov}, {{\v S}ime{\v c}kov{\'a}}, {Tribble}, \&
  {Vincour}}]{Mukhamedzhanov03}
{Mukhamedzhanov}, A.~M., {B{\'e}m}, P., {Brown}, B.~A., {et~al.} 2003, \prc,
  67, 065804

\bibitem[{{Palacios} {et~al.}(2006){Palacios}, {Charbonnel}, {Talon}, \&
  {Siess}}]{Palacios06}
{Palacios}, A., {Charbonnel}, C., {Talon}, S., \& {Siess}, L. 2006, \aap, 453,
  261

\bibitem[{{Palacios} {et~al.}(2003){Palacios}, {Talon}, {Charbonnel}, \&
  {Forestini}}]{Palacios03}
{Palacios}, A., {Talon}, S., {Charbonnel}, C., \& {Forestini}, M. 2003, \aap,
  399, 603

\bibitem[{{Pols} {et~al.}(1995){Pols}, {Tout}, {Eggleton}, \&
  {Han}}]{PolsTout1995}
{Pols}, O.~R., {Tout}, C.~A., {Eggleton}, P.~P., \& {Han}, Z. 1995, \mnras,
  274, 964

\bibitem[{{Reimers}(1975)}]{Reimers75}
{Reimers}, D. 1975, Memoires of the Societe Royale des Sciences de Liege, 8,
  369

\bibitem[{{Rosenblum} {et~al.}(2011){Rosenblum}, {Garaud}, {Traxler}, \&
  {Stellmach}}]{RosenblumGaraudetal11}
{Rosenblum}, E., {Garaud}, P., {Traxler}, A., \& {Stellmach}, S. 2011, \apj,
  731, 66

\bibitem[{{Schaerer} {et~al.}(1993){Schaerer}, {Meynet}, {Maeder}, \&
  {Schaller}}]{Schaerer93}
{Schaerer}, D., {Meynet}, G., {Maeder}, A., \& {Schaller}, G. 1993, \aaps, 98,
  523

\bibitem[{{Schaller} {et~al.}(1992){Schaller}, {Schaerer}, {Meynet}, \&
  {Maeder}}]{Schalleretal1992}
{Schaller}, G., {Schaerer}, D., {Meynet}, G., \& {Maeder}, A. 1992, \aaps, 96,
  269

\bibitem[{{Schatz} {et~al.}(1993){Schatz}, {Kaeppeler}, {Koehler}, {Wiescher},
  \& {Trautvetter}}]{Schatz93}
{Schatz}, H., {Kaeppeler}, F., {Koehler}, P.~E., {Wiescher}, M., \&
  {Trautvetter}, H.-P. 1993, \apj, 413, 750

\bibitem[{{Siess}(2009)}]{Siess09}
{Siess}, L. 2009, \aap, 497, 463

\bibitem[{{Siess} {et~al.}(2000){Siess}, {Dufour}, \&
  {Forestini}}]{SiessDufour2000}
{Siess}, L., {Dufour}, E., \& {Forestini}, M. 2000, \aap, 358, 593

\bibitem[{{Smiljanic} {et~al.}(2010){Smiljanic}, {Pasquini}, {Charbonnel}, \&
  {Lagarde}}]{Smiljanic10}
{Smiljanic}, R., {Pasquini}, L., {Charbonnel}, C., \& {Lagarde}, N. 2010, \aap,
  510, A50

\bibitem[{{Stancliffe}(2010)}]{Stancliffe10}
{Stancliffe}, R.~J. 2010, \mnras, 174

\bibitem[{{Stancliffe} {et~al.}(2009){Stancliffe}, {Church}, {Angelou}, \&
  {Lattanzio}}]{Stancliffe09}
{Stancliffe}, R.~J., {Church}, R.~P., {Angelou}, G.~C., \& {Lattanzio}, J.~C.
  2009, \mnras, 396, 2313

\bibitem[{{Stello} {et~al.}(2009){Stello}, {Chaplin}, {Bruntt}, {Creevey},
  {Garc{\'{\i}}a-Hern{\'a}ndez}, {Monteiro}, {Moya}, {Quirion}, {Sousa},
  {Su{\'a}rez}, {Appourchaux}, {Arentoft}, {Ballot}, {Bedding},
  {Christensen-Dalsgaard}, {Elsworth}, {Fletcher}, {Garc{\'{\i}}a}, {Houdek},
  {Jim{\'e}nez-Reyes}, {Kjeldsen}, {New}, {R{\'e}gulo}, {Salabert}, \&
  {Toutain}}]{Stello2009}
{Stello}, D., {Chaplin}, W.~J., {Bruntt}, H., {et~al.} 2009, \apj, 700, 1589

\bibitem[{{Talon} \& {Charbonnel}(1998)}]{TalCha98}
{Talon}, S. \& {Charbonnel}, C. 1998, \aap, 335, 959

\bibitem[{{Talon} \& {Charbonnel}(2003)}]{TalCha03}
{Talon}, S. \& {Charbonnel}, C. 2003, \aap, 405, 1025

\bibitem[{{Talon} \& {Zahn}(1997)}]{TalZah97}
{Talon}, S. \& {Zahn}, J.-P. 1997, \aap, 317, 749

\bibitem[{{Traxler} {et~al.}(2011){Traxler}, {Garaud}, \&
  {Stellmach}}]{Traxleretal11}
{Traxler}, A., {Garaud}, P., \& {Stellmach}, S. 2011, \apjl, 728, L29

\bibitem[{{Ulrich}(1971)}]{Ulrich71}
{Ulrich}, R.~K. 1971, \apj, 168, 57

\bibitem[{{Ulrich}(1972)}]{Ulrich72}
{Ulrich}, R.~K. 1972, \apj, 172, 165

\bibitem[{{Ulrich}(1986)}]{Ulrich1986}
{Ulrich}, R.~K. 1986, \apjl, 306, L37

\bibitem[{{Uttenthaler} \& {Lebzelter}(2010)}]{UttLeb10}
{Uttenthaler}, S. \& {Lebzelter}, T. 2010, \aap, 510, A62

\bibitem[{{Vassiliadis} \& {Wood}(1993)}]{VaWo93}
{Vassiliadis}, E. \& {Wood}, P.~R. 1993, \apj, 413, 641

\bibitem[{{Wagoner}(1969)}]{Wagoner69}
{Wagoner}, R.~V. 1969, \apjs, 18, 247

\bibitem[{{White} {et~al.}(2011){White}, {Bedding}, {Stello}, {Appourchaux},
  {Ballot}, {Benomar}, {Bonanno}, {Broomhall}, {Campante}, {Chaplin},
  {Christensen-Dalsgaard}, {Corsaro}, {Do{\u g}an}, {Elsworth}, {Fletcher},
  {Garc{\'{\i}}a}, {Gaulme}, {Handberg}, {Hekker}, {Huber}, {Karoff},
  {Kjeldsen}, {Mathur}, {Mosser}, {Monteiro}, {R{\'e}gulo}, {Salabert}, {Silva
  Aguirre}, {Thompson}, {Verner}, {Morris}, {Sanderfer}, \&
  {Seader}}]{White2011}
{White}, T.~R., {Bedding}, T.~R., {Stello}, D., {et~al.} 2011, \apjl, 742, L3

\bibitem[{{Wiescher} {et~al.}(1990){Wiescher}, {Gorres}, \&
  {Thielemann}}]{Wiescheretal90}
{Wiescher}, M., {Gorres}, J., \& {Thielemann}, F. 1990, \apj, 363, 340

\bibitem[{{Woosley} {et~al.}(1978){Woosley}, {Fowler}, {Holmes}, \&
  {Zimmerman}}]{Woosley78}
{Woosley}, S.~E., {Fowler}, W.~A., {Holmes}, J.~A., \& {Zimmerman}, B.~A. 1978,
  Atomic Data and Nuclear Data Tables, 22, 371

\bibitem[{{Yi} {et~al.}(2003){Yi}, {Kim}, \& {Demarque}}]{Yi03}
{Yi}, S.~K., {Kim}, Y.-C., \& {Demarque}, P. 2003, \apjs, 144, 259

\bibitem[{{Zahn}(1992)}]{Zahn92}
{Zahn}, J.-P. 1992, \aap, 265, 115

\bibitem[{{Zahn} {et~al.}(1997){Zahn}, {Talon}, \& {Matias}}]{Zahnetal97}
{Zahn}, J.-P., {Talon}, S., \& {Matias}, J. 1997, \aap, 322, 320

\end{thebibliography}

\end{document}